\documentclass[aps,prr,twocolumn,superscriptaddress,amsmath,amssymb,floatfix,longbibliography]{revtex4-1}
\usepackage{graphicx}
\usepackage{layout}
\usepackage{natbib}
\usepackage{dcolumn}
\usepackage{braket}
\usepackage{bm}
\usepackage{color}
\usepackage{verbatim}
\usepackage{hyperref}
\usepackage{ulem}

\usepackage[columnwise]{lineno}

\begin{document}

\title{Cooper quartets in interacting hybrid  superconducting systems}

\author{Luca Chirolli}
\affiliation{NEST, Istituto Nanoscienze CNR, and Scuola Normale Superiore, I-56127 Pisa, Italy}
\affiliation{Quantum Research Center, Technology Innovation Institute, Abu Dhabi, UAE}

\author{Alessandro Braggio}
\affiliation{NEST, Istituto Nanoscienze CNR, and Scuola Normale Superiore, I-56127 Pisa, Italy}

\author{Francesco Giazotto}
\affiliation{NEST, Istituto Nanoscienze CNR, and Scuola Normale Superiore, I-56127 Pisa, Italy}

\begin{abstract}
Cooper quartets represent exotic fermion aggregates describing correlated matter at the basis of charge-$4e$ superconductivity and offer a platform for studying four-body interactions, of interest for topologically protected quantum computing, nuclear matter simulations, and more general strongly correlated matter. Focusing on solid-state systems, we show how to quantum design Cooper quartets in a double-dot system coupled to ordinary superconducting leads through the introduction of an attractive interdot interaction. A fundamentally novel, maximally correlated double-dot ground state, in the form of a superposition of vacuum $|0\rangle$ and four-electron state $|4e\rangle$, emerges as a narrow resonance in a many-body quartet correlator that is accompanied by negligible pair correlations and features a rich phenomenology. The system represents an instance of correlated Andreev matter and the results open the way to the exploration of interaction effects in hybrid superconducting devices, and the study of novel correlated states of matter with ingredients available in a quantum solid-state lab.
\end{abstract}

\maketitle

\section{Introduction}

The Cooper instability predicts a state of matter in which electrons pair up and form a condensate showing unique properties such as zero resistivity and perfect diamagnetism. Making a conceptual jump we can ask ourselves if more complex electron aggregates such as Cooper quartets could form and condense, yielding charge-$4e$ superconductivity. Such a highly correlated fermion state belongs to a family of complex fermion states, that are typically the subject of intensive quantum simulations through quantum computing platforms \cite{feynman1982simulating,lloyd1996universal,abrams1997simulation,aspuru-guzik2005simulated,georgescu2014quantum}, and its isolation could prove useful for topological quantum computation \cite{kitaev2003fault-tolerant}, simulation of nuclear systems \cite{ropke1998four-particle,funaki2008alpha-particle,schuck2013alpha-particle} and quantum gravity \cite{volovik2024fermionic}, and it could offer great insight in the study of four-body and many-body interactions \cite{mizel2004three,peng2009quantum,dai2017four-body,kumar2020large-scale,zhang2022synthesizing}.  In spin 1/2 systems, Pauli's exclusion principle forbids local multiparticle aggregates beyond pairs. Purely four-fermion instabilities have been predicted in higher spin systems \cite{wu2005competing}, or in systems showing an additional quantum number, such as $\alpha$-particle in nuclear physics \cite{ropke1998four-particle,funaki2008alpha-particle,schuck2013alpha-particle}. It has been suggested that a charge-$4e$ superconducting state may emerge in the fluctuating state of systems displaying a two-component condensate  \cite{babaev2004superconductor,berg2009charge-4e,herland2010phase,lee2014amperean,fradkin2015colloquium}, such as copper-oxides \cite{liPRL2007two-dimensional,dingPRB2008two-dimensional}, kagome systems \cite{lin2021complex,chen2021roton,zhou2022chern,liPRX2023small,guguchia2023unconventional,yu2023nondegenerate}, twisted bilayer graphene \cite{jian2021charge-4e,fernandes2021charge-4e,liu2023charge-4e},  iron-based superconductors \cite{fernandes2014what}, or doped topological insulators \cite{matano2016spin-rotation,shen2017nematic}. In those systems, novel higher-order topological properties are emerging in fermion-quadrupling states and higher composites \cite{grinenko2021state,babaev2024topological,maccari2022effects,maccari2023prediction}. Effective charge-$4e$ superconducting transport properties are obtained in networks of ordinary Josephson junctions, where transport of individual Cooper pairs is suppressed by destructive interference effects and only two-Cooper pair transport can survive  \cite{doucot2002pairing,doucot2003topological,protopopov2004anomalous,doucot2005protected,rizzi20064e-condensation,ciaccia2024charge-4e}, or in multiterminal Andreev bound state in the dissipative regime \cite{cuevas2007voltage-induced,freyn2011production,jonckheere2013multipair,feinberg2015quartets,yonatan2018nonlocal,huang2022evidence,melo2022multiplet}, yielding a phenomenology similar to the one associated to the proper charge-$4e$ state. 

\begin{figure*}[t]
	\centering
	\includegraphics[width=1.0\textwidth]{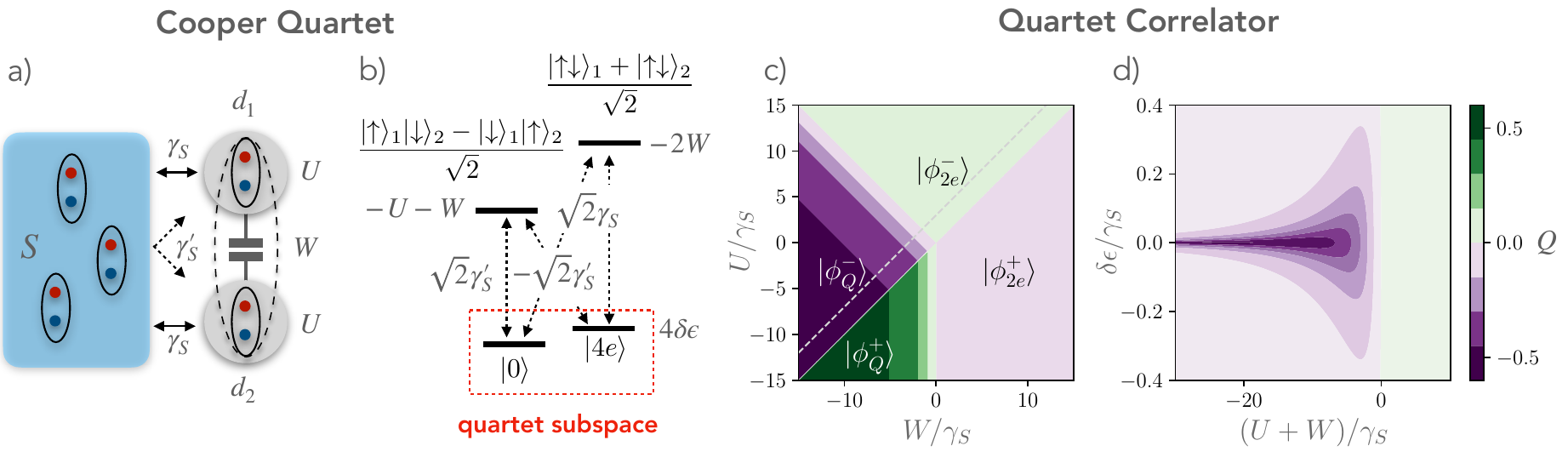}
	\caption{a) Schematics of the double-dot system with onsite interaction $U$, interdot density-density interaction $W$, tunnel-coupled to a superconductor $S$ through rates $\gamma_S$ and $\gamma_S'$. b) Scheme of energy levels and their coupling when the resonance condition is met, $\epsilon=-U/2-W+\delta\epsilon$, up to a small detuning $\delta\epsilon$. c) Phase diagram of the quartet correlator $Q$ at resonance $\epsilon=-U/2-W$ for $\gamma_S'=\gamma_S$: for  $-W>U>0$ the ground state has a value $Q=-1/2$ compatible with a quartet state $|\phi^-_Q\rangle$, whereas for $U<W<0$ the ground state has a value $Q=1/2$ associated to the quartet state $|\phi^+_Q\rangle$. d) Quartet resonance as a function of $\delta \epsilon=\epsilon+U/2+W$ along the dashed gray line in c), where $U-W=3\gamma_S$, showing the dependence of the correlator $Q$ on the detuning of the vacuum and four-electron states.
	\label{Fig1}}
\end{figure*}

In this work, we change the perspective and tailor the conditions for the onset of a Cooper quartet ground state.  We consider a minimal model of a double quantum dot, that allows us to accommodate a four-electron superposition state through an additional orbital binary quantum number beyond spin. We exploit Cooper pairing through proximity coupling the double-dot system to a Bardeen-Cooper-Schrieffer (BCS)  superconductor. We then add a crucial ingredient, that consists in an strong attractive interaction between the quantum dots and that allows us to stabilize the quartet superconducting phase. An attractive interdot interactions has be experimentally realized through a second double quantum dot \cite{hamo2016electron}, or through a transmission line resonator \cite{delbecq2013photon-mediated}, and it has been proposed to emerge by  coupling to a flexural phonon in a suspended carbon nanotube \cite{bhattacharya2021phonon,vigneauPRRultrastrong,moser2013ultrasensitive}. The system manifests a quartet ground state
\begin{equation}\label{Eq:QuartetStates}
|\phi^\pm_{\rm Q}\rangle=\frac{1}{\sqrt 2}(|0\rangle \pm |4e\rangle),
\end{equation}
in which the vacuum state $|0\rangle$ and the four-electron state $|4e\rangle$ appear in superposition of equal weights and that describes the formation of a coherent, albeit minimal, condensate, whose properties are dictated by the charge $4e$. We show that many-body interactions yield a finite quartet correlator,
\begin{eqnarray}
Q&=&\langle d_{1\downarrow}d_{1\uparrow}d_{2\downarrow}d_{2\uparrow}\rangle-\langle d_{1\downarrow}d_{1\uparrow}\rangle\langle d_{2\downarrow}d_{2\uparrow}\rangle\nonumber\\
&-&\langle d_{1\downarrow}d_{2\uparrow}\rangle\langle d_{1\uparrow}d_{2\downarrow} \rangle, 
\end{eqnarray}
with $d_{i\sigma}$ fermionic dot annihilation operators, showing that {\it pure} quartet superconductivity violates the Wick theorem and has a genuine interacting origin. It is important to stress indeed that quartet correlations may be present in a BCS superconductor, but are subdominant with respect to pair correlations. The correlator $Q$ resonantly peaks to its extreme values $\pm 1/2$ when the states $|\phi_Q^\pm\rangle$ are realized, whereas the pair correlator is zero on these states. We study the Josephson currents by attaching extra BCS leads and single out the conditions for a two-Cooper pair current. More generally, the system realizes a correlated Andreev matter beyond the family of multi-terminal Josephson junctions \cite{riwar2016multi-terminal,teshler2023ground,ohnmacht2023quartet,zalom2024hidden}, that is generated by interactions beyond the standard BCS mean field terms. The quartet ground state delocalized in the two quantum dots shows a strong non-local coherent character and we identify a novel non-local phase response in a three-terminal setup as a signature of the correlated nature of the ground state.

These findings open the way to the exploration of interaction-based superconductors and non-local phase coherence of multiterminal setups can be used to better investigate those systems \cite{coraiola2023phase-engineering,pankratova2020multiterminal,matsuo2022observation}. Applications in parity-protected quantum computing schemes, such as those based on two-Cooper pair transport  \cite{blatter2001design,Ioffe2002topologically,ioffe2002possible,brooks2013protected,doucot2012physical,brosco2024superconducting,coppo2024flux-tunable}, and more generally in the simulation of novel phases of matter constituted by exotic electron complexes can be envisioned using simple tools and ingredients available in a solid-state quantum lab.

\section{The model}

We consider the double quantum dot system schematized in Fig.~\ref{Fig1}a), with the two dots labeled with $i=1,2$, tunnel coupled to a common superconducting lead $S$ at the left of the system. We describe their Hamiltonian $H_0$ through single-electron gate-tunable levels with energy $\epsilon_i$,  an intradot (generally repulsive) Hubbard term with strength $U$, and an interdot density-density interaction with strength $W$. To theoretically investigate the rich phase diagram of the system we allow interactions to be attractive, and refer to possible underlaying mechanisms that have been theoretically and experimentally investigated  \cite{hamo2016electron,delbecq2013photon-mediated,bhattacharya2021phonon,vigneauPRRultrastrong,moser2013ultrasensitive}. The double-dot Hamiltonian reads
\begin{eqnarray}
    H_0&=&\sum_{i,\sigma}\epsilon_{i}n_{i\sigma}+U\sum_in_{i\uparrow}n_{i\downarrow}+Wn_1n_2,
\end{eqnarray}
with $n_i=n_{i\uparrow}+n_{i\downarrow}$, $n_{i\sigma}=d^\dag_{i\sigma}d_{i\sigma}$, and $d_{i\sigma}$ dot fermionic  annihilation operators. We include tunnel coupling to the superconducting lead $S$ as a proximity effect, for which the quantum dots develop a finite pairing amplitude to form Cooper pairs, either locally on each dot or in a delocalized way, between the two dots, so that the pairing Hamiltonian reads
\begin{equation}
H_p=\gamma_S\sum_{i}d^\dag_{i\uparrow}d^\dag_{i\downarrow}+\gamma'_S(d^\dag_{1\uparrow}d^\dag_{2\downarrow}-{d^\dag_{1\downarrow}d^\dag_{2,\uparrow}})+{\rm H.c.}.
\end{equation}
This model has been thoroughly studied in the infinite superconducting gap limit ($\Delta\to \infty$), where scattering off the superconducting lead becomes elastic and integration away of the leads produces effective pairing amplitude, $\gamma_S=2\pi\nu_Ft^2_{DS}$ and $\gamma'_S=\frac{1}{2}\gamma_S e^{-\delta r/\xi}\sin(k_F\delta r)/(k_F\delta r)$. These coincide with the rates of local and crossed Andreev reflection, in which a Cooper pair can either tunnel to one of the two dots or split between the two dots, respectively, with $\nu_F$ the density of states at the Fermi energy, $t_{DS}$ the dot-superconductor tunneling amplitude, assumed equal for the two dots, $\delta r$ the interdot distance, $k_F$ the Fermi momentum, and $\xi$ the  coherence length of the superconductor.

The full Hamiltonian is 
\begin{equation}\label{Eq:HamiltonianFull}
H=H_0+H_p,
\end{equation}
and we focus the analysis on the even parity sector, where the four electron operator $d_{1\downarrow}d_{1\uparrow}d_{2\downarrow}d_{2\uparrow}$ can have non-zero matrix element. We further restrict ourselves to the subspace spanned by the following four states: the vacuum $|0\rangle$, the four electron state $|4e\rangle\equiv|{\uparrow\downarrow}\rangle_1|{\uparrow\downarrow}\rangle_2$, and the two doubly occupied states $|\phi_{2e}^+\rangle\equiv(|{\uparrow\downarrow}\rangle_1+|{\uparrow\downarrow}\rangle_2)/\sqrt{2}$ and $|\phi_{2e}^-\rangle\equiv(|{\uparrow}\rangle_1|{\downarrow}\rangle_2-|{\downarrow}\rangle_1|{\uparrow}\rangle_2)/\sqrt{2}$.  In this subspace the Hamiltonian reads
\begin{equation}\label{Eq:H4x4}
H_e=\left(\begin{array}{cccc}
0 & 0 &  \sqrt{2}\gamma_S &  \sqrt{2}\gamma'_S\\
0 & 4\epsilon+2U+4W & \sqrt{2}\gamma_S & -\sqrt{2}\gamma_S'\\
\sqrt{2}\gamma_S & \sqrt{2}\gamma_S & 2\epsilon+U & 0\\
\sqrt{2}\gamma'_S & -\sqrt{2}\gamma'_S & 0 &  2\epsilon+W\\
\end{array}\right).
\end{equation}

Quartet correlations emerge specifically around the resonance condition 
\begin{equation}
4\epsilon+2U+4W=0,
\end{equation}
for which the vacuum $|0\rangle$ and the four-fold occupied state $|4e\rangle$ become degenerate. In addition, the Hamiltonian decouples in two subspaces spanned by  $\{|\phi^+_Q\rangle,|\phi^+_{2e}\rangle\}$ and $\{|\phi^-_{Q}\rangle,|\phi^-_{2e}\rangle\}$. To select the doublet $|0\rangle$ and $|4e\rangle$ as a host for a quartet ground state, we require all other states to have higher energy. By realistically assuming a repulsive onsite interaction $U>0$, a relevant condition is met when the quantum dots experience an attractive density-density interaction that is stronger than the onsite repulsion, $-W>U>0$. Close to resonance, $\epsilon=-U/2-W+\delta\epsilon$, with $\delta\epsilon$ a weak detuning from resonance, and for $-2W>0$, $-U-W>0$, and $\gamma_S,\gamma_S'\ll |U+W|, U,|W|$, the vacuum and the four-electron states couple at first order with the higher energy states $|\phi_{2e}^+\rangle$
and $|\phi_{2e}^-\rangle$
, as schematized in Fig.~\ref{Fig1}b), and result in the low energy states 
\begin{eqnarray}\label{Eq:LEstates}
|\bar{0}\rangle=|0\rangle+\frac{\sqrt{2}\gamma_S}{2W-2\delta\epsilon}|\phi^+_{2e}\rangle+\frac{\sqrt{2}\gamma'_S}{U+W-2\delta\epsilon}|\phi^-_{2e}\rangle,\\
|\bar{4e}\rangle=|4e\rangle+\frac{\sqrt{2}\gamma_S}{2W-2\delta\epsilon}|\phi^+_{2e}\rangle-\frac{\sqrt{2}\gamma'_S}{U+W-2\delta\epsilon}|\phi^-_{2e}\rangle.
\end{eqnarray}
We then project the Hamiltonian on the low energy states, where it assumes a simple form 
\begin{equation}
h=-2\delta\epsilon(|\bar{0}\rangle\langle{0}|-|\bar{4e}\rangle\langle\bar{4e}|)+\Gamma(|\bar{0}\rangle\langle\bar{4e}|+{\rm H.c.}),
\end{equation}
with the coupling matrix element 
\begin{equation}
\Gamma=-\frac{2(\gamma_S')^2}{U+W}+\frac{\gamma_S^2}{W}.
\end{equation}

The phase diagram of the system is shown in Fig.~\ref{Fig1}c), where we plot the ground state value of correlator $Q$ at resonance $\epsilon=-U/2-W$. For completeness, we present results for positive and negative values of $U$ and $W$. The correlator can take generic complex values satisfying $|Q|\leq 1/2$ and, for real $\gamma_S,\gamma_S'$, it is real (see Appendix \ref{Appendix}). In the region $U>W$ the ground state belongs to the subspace spanned by $\{|\phi^-_Q\rangle,|\phi^-_{2e}\rangle\}$ and the correlator evolves from $Q=-1/2$, when $U+W<0$ and the ground state has a strong $|\phi^-_Q\rangle$ quartet character, to $Q\sim (\gamma_S')^2/(U+W)^2$ when $U+W>0$, where the ground state has mostly a two-electron state character $|\phi^-_{2e}\rangle$.  In the region $U<W$ of the phase diagram Fig.~\ref{Fig1}c) the ground state belongs to the sector spanned by $\{|\phi^+_Q\rangle,|\phi^+_{2e}\rangle\}$, and for the fully attractive case $W,U<0$ a quartet ground state $|\phi^+_Q\rangle$ is achieved for which $Q= 1/2$. Moving away from resonance, for $\epsilon=-U/2-W+\delta\epsilon$, the value of the correlator gives a Breit-Wigner resonance $Q\simeq -\Gamma/(2\sqrt{\Gamma^2+4\delta\epsilon^2})$ with linewidth $\Gamma$ for $U+W<0$, as shown in Fig.~\ref{Fig1}d), and away from resonance the quartet correlations rapidly decaying. Importantly, when the ground state belongs to the sector spanned by $\{|\bar{0}\rangle,|\bar{4e}\rangle\}$ the pair correlator is on order $\gamma_S/W$ or $\gamma_S'/(U+W)$.

\section{Correlated Andreev matter}

\begin{figure*}[t]
	\centering
	\includegraphics[width=1.0\textwidth]{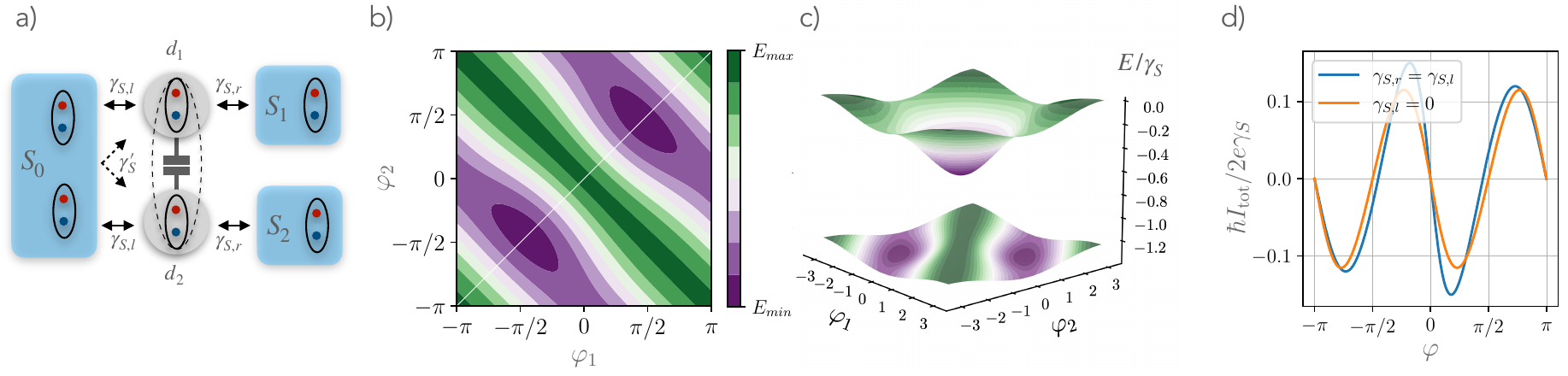}
	\caption{a) Schematics of the three-terminal structure composed by two quantum dots, 1 and 2, coupled to a common superconducting lead $S_0$ 	at the left and to superconducting leads, $S_1$ and $S_2$, at the right. The latter are biased at phases $\varphi_1$ and $\varphi_2$ with respect to 	$S_0$, respectively.  b) Contour plot of the ground state energy of the system in a) with respect to the phases $\varphi_1$ and $\varphi_2$, showing two minima at approximately 
	$\varphi_1=\varphi_2\simeq \pm \pi/2$, for  $\gamma_S'=\gamma_{S,l}=\gamma_{S,r}=\gamma_S$, $\epsilon=-U/2-W$, $U/\gamma_S=10$ and 	$W/\gamma_S=-12$. c)  Plot of the ground state and the first excited states as a function of the phases $\varphi_1$ and $\varphi_2$, representing a correlated Andreev matter and realizing a quartet Andreev qubit. d) Total current $I_1+I_2$ as a function of $\varphi_1=\varphi_2=\varphi$ along the white line in panel c), shown for the case 
	$\gamma_{S,l}=\gamma_{S,r}$, as in b) and c), and for $\gamma_{S,l}=0$ for which the current is exactly $\pi$-periodic.}
	\label{Fig2}
\end{figure*}

It becomes natural at this point to search for a direct consequence of having a quartet ground state and a smoking gun observable that can witness its presence. The most intuitive property that a quartet ground state is expected to show is a quartet dissipationless current. However, since the leads are BCS superconductors, the fundamental carriers are Cooper-pairs and a  Josephson current will in general flow through  higher  energy two-electron states, at second order in the rates $\gamma_S,\gamma_S'$, resulting in a standard $2\pi$-periodic current-phase relation, despite the fact that the double-dot system is in a quartet ground state. In particular, in a specular two-terminal configuration with an additional superconducting lead on the right side, coupled with same $\gamma_S, \gamma_S'$, the rates all acquire a factor $2\cos(\varphi/2)$ and the resulting current is $2\pi$ periodic, reflecting the fact that the system can load and discharge individual Cooper pairs via higher energy states through multiple processes involving local and non-local Andreev reflection. 

Nevertheless, the system features a strongly correlated ground state and we expect that under proper conditions it can mediate a non-trivial two-Cooper pair current, that appears as a $\pi$-periodic Josephson current. For this to occur, a peculiar interplay between local and non-local Andreev processes needs to be arranged. More generally, the delocalized nature of the quartet ground state suggests that the system can manifest a correlated behavior in a multi-terminal configuration and a more general class of correlated multi-terminal Andreev matter can emerge.

To generalize the description we focus on a three-terminal structure constituted by a common lead at the left, relabeled lead 0, and two additional leads 1 and 2 each coupled to the corresponding quantum dot, as schematized in Fig.~\ref{Fig2}a), and study the supercurrents in the system. A proper description requires enlarging the basis and separately consider four two-electron states together with the vacuum $|0\rangle$ and the four-electron state $|4e\rangle$, so that the full basis reads $\{|0\rangle,|{\uparrow\downarrow}\rangle_1$, $|{\uparrow\downarrow}\rangle_2$, $|{\uparrow}\rangle_1|{\downarrow}\rangle_2$, $|{\downarrow}\rangle_1|{\uparrow}\rangle_2,|4e\rangle\}$. Proximity-induced local pairing in quantum dot $i=1,2$ can originate either from lead 0 or from the corresponding $i$-lead at the right which is biased at phase $\varphi_i$ with respect to the reference lead 0, as schematized in Fig.~\ref{Fig2}a). We assume different local rates at the left $\gamma_{S,l}$ and at the right $\gamma_{S,r}$ leads, equal for the two dots. This way, local pairing in dot $i$ becomes controlled by $\gamma_{S,i}=\gamma_{S,l}+\gamma_{S,r}e^{i\varphi_i}$.  In turn, non-local pairing $\gamma_S'$ can only result from Cooper pair splitting from lead 0 and it is not sensitive to any phase differences.

The low energy states are modified by the complex rates and now read
\begin{eqnarray}
|\bar{0}\rangle&=&|0\rangle+\sum_i\frac{\gamma_{S,i}}{2W-2\delta\epsilon}|{\uparrow\downarrow}\rangle_i+\frac{\sqrt{2}\gamma_S'}{U+W-2\delta\epsilon}|\phi^-_{2e}\rangle,\nonumber\\
|\bar{4e}\rangle&=&|4e\rangle+\frac{\gamma^*_{S,2}|{\uparrow\downarrow}\rangle_1+\gamma^*_{S,1}|{\uparrow\downarrow}\rangle_2}{2W-2\delta\epsilon}-\frac{\sqrt{2}\gamma_S'}{U+W-2\delta\epsilon}|\phi^-_{2e}\rangle,\nonumber
\end{eqnarray}
and the Hamiltonian projected on the low-energy manifold reads 
\begin{equation}\label{Eq:HamAndreev2x2}
h=\frac{\gamma_{S,l}\gamma_{S,r}}{W}(\cos(\varphi_1)+\cos(\varphi_2))\openone+\left(
\begin{array}{cc}
-2\delta \epsilon &  \Gamma\\
\Gamma^* & 2\delta\epsilon
\end{array}
\right),
\end{equation}
with $\openone$ a $2\times 2$ identity in the low energy manifold and $\Gamma$ a generalized complex matrix element between vacuum and four-electron states, that at $\delta\epsilon=0$ reads 
\begin{equation}
\Gamma(\varphi_1,\varphi_2)=-\frac{2(\gamma_S')^2}{U+W}+\frac{1}{W}(\gamma_{S,l}+\gamma_{S,r}e^{i\varphi_1})(\gamma_{S,l}+\gamma_{S,r}e^{i\varphi_2}). 
\end{equation}
Its argument $\Theta(\varphi_1,\varphi_2)={\rm arg}\left[\Gamma\right]$ represents the phase of the superposition between vacuum and four-electron state in the ground state, $|\Psi_{\rm GS}\rangle=\frac{1}{\sqrt 2}(|\bar{0}\rangle-e^{i\Theta}|\bar{4e}\rangle)$, and can be tuned through the phases $\varphi_{1}$ and $\varphi_2$.

The ground state energy as a function of the two phase differences $\varphi_1,\varphi_2$ is shown in Fig.~\ref{Fig2}b) and it clearly shows two minima at $\varphi_1=\varphi_2\simeq \pm \pi/2$ along the $\varphi_1=\varphi_2$ line, and a saddle point at $\varphi_1=\varphi_2=0$, manifesting the strongly correlated character of the ground state. In particular, we notice that if the leads $S_1$ and $S_2$ are left floating, so that the system adjusts in one minimum of the ground state energy,  the latter breaks time-reversal symmetry.

The Hamiltonian Eq.~\eqref{Eq:HamAndreev2x2} well captures also the first excited state, whose exact landscape is shown in Fig.~\ref{Fig2}c) together with the ground state at zero detuning $\delta\epsilon$. The system realizes a quartet Andreev qubit that  represents an instance of correlated Andreev matter beyond the family of multi-terminal Josephson junctions \cite{riwar2016multi-terminal,teshler2023ground,ohnmacht2023quartet,zalom2024hidden}, and possibly featuring topological properties that will be discussed in future works.

\section{Dissipationless transport}

\begin{figure*}[t]
	\centering
	\includegraphics[width=1.0\textwidth]{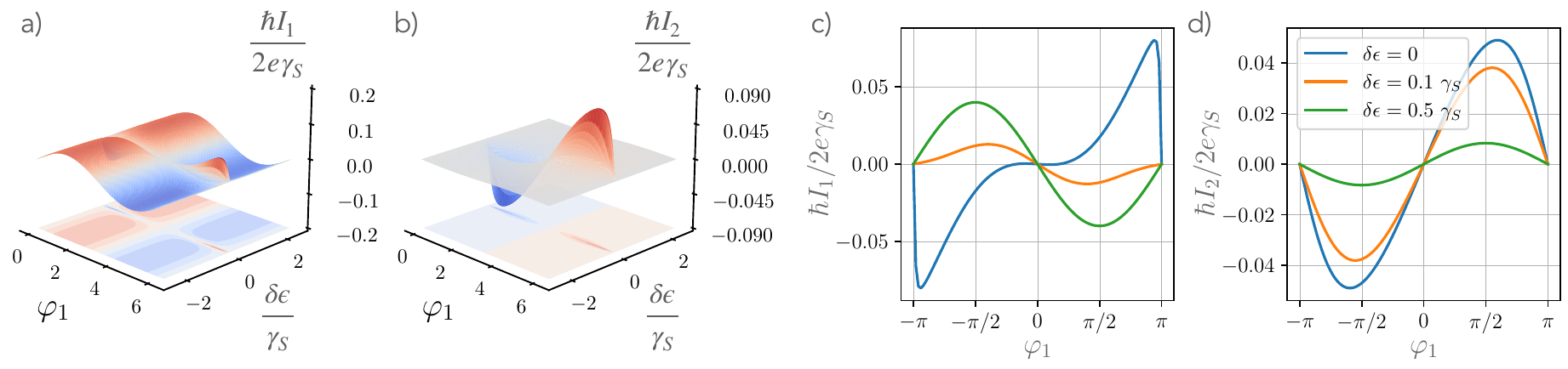}
	\caption{a) Current $I_1(\varphi_1,\varphi_2=0)$ and b) $I_2(\varphi_1,\varphi_2=0)$ as a function of the phase $\varphi_1$ and the detuning $\delta\epsilon$, obtained by keeping $\varphi_2=0$. c) Cuts of $I_1(\varphi_1,\varphi_2=0)$ in a) at three different values of $\delta\epsilon$ shown in the legend of d). d) Cuts of $I_2(\varphi_1,\varphi_2=0)$ at the values of $\delta\epsilon$ specified in the legend. The parameters for the plots are the same as in Fig.~\ref{Fig2}.}
	\label{Fig3}
\end{figure*}

We now analyze the dissipationless transport properties of the system. The ground-state current through dot $i$ features two contributions: an ordinary Josephson term describing current between leads $i$ and 0, and a second term that depends on both $\varphi_1$ and $\varphi_2$,
\begin{equation}\label{Eq:currents}
I_i=-\frac{2e}{\hbar}\frac{\gamma_{S,l}\gamma_{S,r}}{W}\sin(\varphi_i)-\frac{2e}{\hbar}\frac{\partial}{\partial \varphi_i}\left|\Gamma(\varphi_1,\varphi_2)\right|.
\end{equation}

We first look for a two-Cooper pair current in a two-terminal configuration by joining contact 1 and 2 in a unique drain by setting $\varphi_1=\varphi_2=\varphi$ and study the total current $I=I_1+I_2$. An ideal two-Cooper pair current flows in the case $\gamma_{S,l}=0$. This is easily understood as follows. Starting in the quartet ground state a Cooper pair coming from the left lead can only split between the two dots, by virtue of the condition $\gamma_{S,l}=0$. Such a state is current-blocked by the superconducting gap in the leads 1 and 2, that have no single-particle states in the spectrum (especially in the $\Delta\to\infty$ limit). A second Cooper pair coming from the left lead can again only split and result in a four-electron state in the double-dot. The latter is now unblocked and gives rise to an ideal two-Cooper pair current, as shown in Fig.~\ref{Fig2}d). This intuitive picture is confirmed by the form of the ground state energy that, setting $\gamma_{S,r}=\gamma_S$, reads
\begin{equation}\label{Eq:SM-Egs}
    E_{\rm GS}=-E_0\sqrt{1+\tau\sin^2(\varphi)},
\end{equation}
with $E_0=\left|-\frac{2(\gamma_S')^2}{U+W}+\frac{\gamma_S^2}{W}\right|$ and $\tau=\frac{8(\gamma_S\gamma_S')^2}{E_0^2W(U+W)}$, that is manifestly $\pi$-periodic. In addition, for $\tau>0$ the minima of the ground state energy Eq.~(\ref{Eq:SM-Egs}) are at $\varphi=\pm \pi/2$, so that the current at small finite bias flows in the direction opposite to the phase bias, in an effective $\pi$-junction behavior, as shown in Fig.~\ref{Fig2}d).

\begin{figure*}[t]
	\centering
	\includegraphics[width=1.0\textwidth]{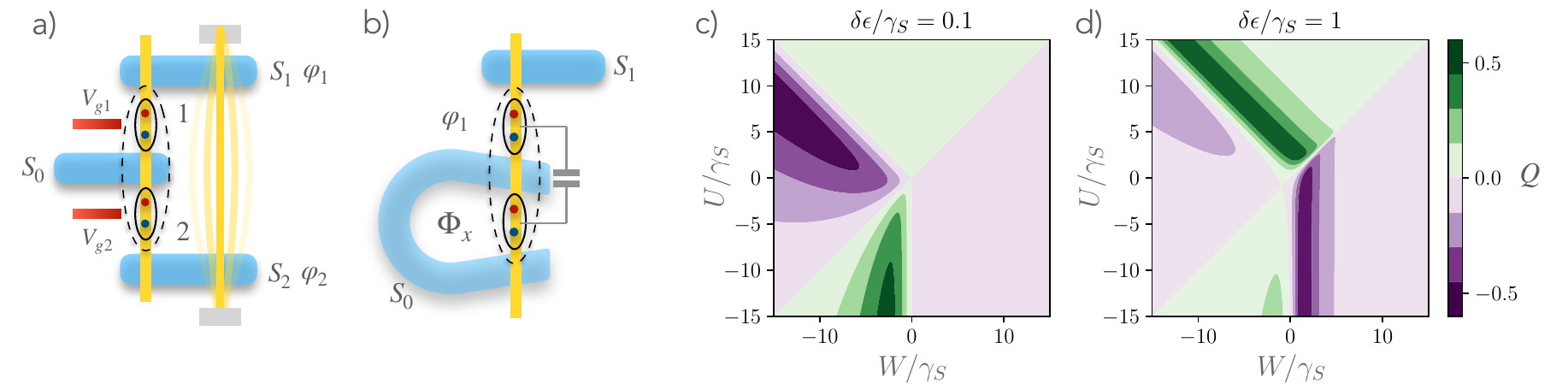}
	\caption{a) Schematics of a possible implementation of the system. The double quantum dot is formed on a carbon nanotube in tunnel contact with superconducting leads $S_0$, $S_1$, and $S_2$. Gates $V_{g1}$ and $V_{g2}$ allow for alignment of the dots levels. A second carbon nanotube, on the right side is suspended between contacts $S_1$ and $S_2$ but not in tunnel contact. Its vibrating lowest energy flexural mode couples to the charge on the quantum dots and provided a mechanism for strong interdot density-density interaction. b) Modified setup that allows the pinning of the phase $\varphi_2=2e\Phi_x/h$ and the resulting current $I_1$ shown in Fig.~\ref{Fig3}. c) Quartet correlator $Q$ away from resonance, with $\delta\epsilon/\gamma_S=0.1$ and d) $\delta\epsilon/\gamma_S=1$, with $\gamma_S'=\gamma_S$.   }
	\label{Fig4}
\end{figure*}

For the realistic case of $\gamma_{S,l}=\gamma_{S,r}=\gamma_S$, the different terms compete and a $\pi$-periodic current can only arise approximately, in that the contribution coming from $|\Gamma|$ must cancel the ordinary $2\pi$-periodic pair contribution of the first term Eq.~\eqref{Eq:currents}. This can be achieved by  promoting the relative importance of the non-local process with respect to the local one in the expression  of $\Gamma$ by requiring $|2(\gamma_S')^2/(U+W)|\gg |\gamma_S^2/W|$. Crucial is the opposite sign between the terms that originates  from the singlet expression of the non-local Cooper pair tunneling and yields a destructive interference effect. By setting $W=-U-\delta W$, $\delta W\ll U$, $\gamma_S'=\gamma_S$, and expanding the current at lowest orders in $\delta W/U$ we obtain
\begin{equation}
I=-\frac{4e\gamma^2_S}{\hbar U}\left[\sin(2\varphi)-{\cal O}\left(\delta W/U\right)\right].
\end{equation}
The resulting current is shown in Fig.~\ref{Fig2}d) and shows a dominant $-\sin(2\varphi)$ component. Importantly, we obtain such a condition without recurrence to any flux interferometric methods \cite{doucot2012physical}.

\section{Non-local phase response}

The interacting nature of the quartet correlator suggests us that another possible clear signature of the quartet ground state can appear in the non-local response of the system. A non-local response appears by noticing that a given phase difference established between two given terminals, say 0 and 2, affects the current through the third terminal in a non-trivial way. If we close lead 2 on lead 0 and enclose a flux $\Phi_x$ (as schematized in Fig.~\ref{Fig4}b)) we effectively pin the phase $\varphi_2=2e\Phi_x/\hbar$, so that we can selectively choose an horizontal cut in the energy landscape Fig.~\ref{Fig2}c). The resulting current $I_1$ is shown in Fig.~\ref{Fig3}a) as a function of the phase $\varphi_1$ and the detuning $\delta\epsilon$ and for three specifics values of $\delta\epsilon$ in Fig.~\ref{Fig3}c): the application of a phase bias $\varphi_1$ between terminal 1 and 0 at constant $\varphi_2=0$ yields a sinusoidal current $I_1(\varphi_1)$, away from the quartet resonance condition, that sharply evolves in a current of opposite sign at resonance $\delta\epsilon=0$, in a sort of effective $\pi$-junction behaviour. At the same time, at resonance a current $I_2(\varphi_1)$ starts to flow in the ring at zero external flux, with the same negative sign as $I_1$, as shown in Fig.~\ref{Fig3}b) and d).

This behavior constitutes a clear signature of the quartet Andreev process for this simple setup, that becomes active only at resonance, and it is understood as follows: away from the resonance the ground state has a weak quartet component and an ordinary dissipationless current flows from terminal 2 to terminal 0, following the phase bias. As resonance is approached, the system develops a ground state with a strong quartet component, that  necessarily yields currents in both terminal 1 and 2, 
that goes in the opposite direction to the phase bias, as dictated by the interference between the phase dependent and the phase independent term that constitute $\Gamma$.

It is worth to remind that the emergence of a $\pi$-junction behavior typically requires engineering complex structures, that involve ferromagnetic systems \cite{golubov2001coupling},  unconventional order parameters \cite{tsuei2000pairing}, non-equilibrium effects \cite{baselmans1999reversing}, or weak Kondo correlations \cite{cleuziou2006carbon}. Furthermore, the $I_2$ current is anomalous in that it flows in a zero phase bias condition in absence of any time-reversal breaking mechanism other than the phase $\varphi_1$. Typically anomalous Josephson effect requires strong spin-orbit interaction and a Zeeman field \cite{szombati2016josephson}, conditions that are not verified in our case.  

\section{Possible experimental setup}

Here, we provide a possible experimental implementation of a system that could host the attractive double quantum dot and several other possibility have been explored in literature \cite{hamo2016electron,delbecq2013photon-mediated,bhattacharya2021phonon,vigneauPRRultrastrong,moser2013ultrasensitive}. The system is shown in Fig.~\ref{Fig4}a) and it is composed by two carbon nanotubes: the left one can be placed to be in tunnel contact with the three superconducting contacts $S_0$, $S_1$, and $S_2$, with $S_0$ at the center, and the right one suspended between contacts $S_1$ and $S_2$, but not galvanically coupled, and in close proximity with the left one. This way, the contacts define two dots and the central one can inject Cooper pairs locally and non-locally. The charge on the two dots couples to the charge on the suspended nanotube via Coulomb interaction in the vacuum. The latter is not screened and the coupling is thus given by the long range Coulomb interaction $\propto 1/r^2$. The bending of the suspended nanotube depends on the total charge on the double quantum dots and the minimal Hamiltonian consisting of a single nanotube flexural mode $a$ reads
\begin{equation}
H=\omega_0 a^\dag a+g_0(a+a^\dag)(n_1+n_2).
\end{equation}
with $a$ denoting he bosonic annihilation operator of the flexural mode. The Hamiltonian is quadratic in the bosonic modes, so that it can be rewritten as
$H=\omega_0 A^\dag A-(g_0^2/\omega_0)(n_1+n_2)^2$, where the new displaced bosonic mode reads $A=a+(g_0/\omega_0)(n_1+n_2)$. The ground state will be given by the vacuum of $A$, so that the correction of the energy results in the effective interaction for the two dots
\begin{equation}
H_{\rm int}=U(n_{1\uparrow}n_{1\downarrow}+n_{2\uparrow}n_{2\downarrow})+Wn_1n_2,
\end{equation}
with $U=U_0-\frac{2g_0^2}{\omega_0}$ and $W=W_0-\frac{2g_0^2}{\omega_0}$, with $U_0>0$ and $W_0>0$ the bare onsite repulsion and interdot density-density interaction strengths. The crucial point is that the bare frequency of the flexural phonon can be very low compared to the coupling to the electronic density \cite{mahan2004}, so that a strong renormalization of the bare interaction is possible. For bare energy scales $U_0\simeq 1~{\rm THz}$, $\omega_0\simeq1~{\rm MHz}$, and $g_0\simeq 1~{\rm GHz}$, values of the coupling constant $0.25< g_0^2/(U_0\omega_0) <0.35$ can be realized \cite{bhattacharya2021phonon,vigneauPRRultrastrong,moser2013ultrasensitive}.

\section{Conclusions}

We have demonstrated how appropriate tuning of electron-electron interaction in a double quantum dot system opens the possibility to engineer a quartet superconductor. The system presented shows a great host of counter-intuitive properties and opens unexplored paths in the simulation of correlated states of matter and the study of interactions effects superconductors. As a further development in the context of transport in hybrid normal-quantum-dot-superconducting devices we foresee the possibility of a quartet Andreev reflection, something that may make the quartet ground state nature evident and that will be addressed in future works.

The possibility to experimentally achieve the isolation of a quartet ground state relies on the ability to engineer an attractive density-density interaction. The latter has been mediated by a second double quantum dot capacitively coupled to the system  \cite{hamo2016electron}, or through a transmission line resonator \cite{delbecq2013photon-mediated}. In addition, we have presented a simple possible setup based on a suspended carbon nanotube as a possible effective mediator of the attractive interdot interaction.

The quartet resonance is fragile to detuning $\delta\epsilon$, as shown in Fig.~\ref{Fig1}d), but also to temperature, in that the correlator is washed away for temperature $T$ larger than $|\Gamma|$, that separates the ground state $|\phi^-_Q\rangle$ from the excited state $|\phi^+_Q\rangle$. Accounting for the latter results in $Q\simeq -(1-e^{-|\Gamma|/T})/2$. Nevertheless, if the hierarchy of scales $T,\delta\epsilon<|\Gamma|<\gamma_S,\gamma_S'\ll U,|W|< \Delta$, is respected the picture holds. Assuming $\gamma_S,\gamma_S'\simeq 20~{\rm GHz}$ \cite{biercuk2005gate-defined,hofstetter2009cooper,herrmann2010carbon}, we expect that temperatures of the order of $25$ mK should be enough to stabilize a quartet ground state.  Furthermore, relaxing the infinite gap condition does not change drastically the picture, so long as the entire double-dot spectrum falls within the superconducting gap $\Delta$. 

In the finalization of the work, we became aware of recent studies concerning quartet superconductivity in a two-orbital Hubbard model with similar results \cite{soldini2024charge4e}.

{\bf Acknowledgment.---} We acknowledge useful discussions with W. Belzig, L. Amico. We acknowledge the EU’s Horizon 2020 Research and Innovation Framework Programme under Grant No. 964398 (SUPERGATE), No. 101057977 (SPECTRUM), and the PNRR MUR project PE0000023-NQSTI for partial financial support. The work has been also supported by the Fondazione Cariplo on the project N. 2023-2594. A.B. acknowledges MUR-PRIN 2022 - Grant No. 2022B9P8LN-(PE3)-Project NEThEQS “Non-equilibrium coherent thermal effects in quantum systems” in PNRR Mission
4-Component 2-Investment 1.1 “Fondo per il Programma Nazionale di Ricerca e Progetti di Rilevante Interesse Nazionale (PRIN)” funded by the European Union-Next Generation EU, the Royal Society through the International Exchanges between the UK and Italy (Grants No.
IEC R2 192166) and CNR project QTHERMONANO.

\appendix

\section{Correlator}
\label{Appendix}

When the Wick theorem applies, the expectation value of any product of operators on the ground state can be decomposed in all possible two-operator contractions. In the case of quartet superconductivity, the Wick theorem does not apply and the correlator itself has been defined as cleaned by the two-point contractions, so that when the state is a Slater determinant the correlator is zero. In the definition we have omitted the equal-spin two-point contractions, that are zero in our case. 

In general the Hamiltonian Eq.~\eqref{Eq:HamiltonianFull} gives rise to a complex matrix representation. However, for the case of a single superconducting contact or for no phase differences between the terminals in the case of a multi-terminal structure, it is possible to choose a gauge in which the Hamiltonian is real and so are the eigenstates. Assuming a generic state
\begin{eqnarray}
|\psi\rangle&=&\cos(\theta_1)|0\rangle+\sin(\theta_1)\cos(\theta_2)|\phi^+_{2e}\rangle\nonumber\\
&+&\sin(\theta_1)\sin(\theta_2)\cos(\theta_3)|\phi^-_{2e}\rangle\nonumber\\
&+&\sin(\theta_1)\sin(\theta_2)\sin(\theta_3)|4e\rangle,
\end{eqnarray}
The maximum of $|Q|$ is obtained for the values $\theta_1=\pi/4$ and $\theta_2=\theta_3=\pm \pi/2$, corresponding to the states $|\phi^\pm_Q\rangle$ for which the correlator takes the values $Q=\pm 1/2$. Numerical checks on eight-component even parity states with real coefficients confirm the bounds $-1/2\leq Q\leq 1/2$.

The value of the correlator is therefore very significant of the quartet content of the ground state and for completeness in Fig.~\ref{Fig4}c) and d) we show its value slightly away from exact resonance, for $\delta\epsilon/\gamma_S=0.1,1$ and for $\gamma_S'=\gamma_S$.

\bibliography{quartets-bib}{}

\begin{thebibliography}{81}%
\makeatletter
\providecommand \@ifxundefined [1]{%
 \@ifx{#1\undefined}
}%
\providecommand \@ifnum [1]{%
 \ifnum #1\expandafter \@firstoftwo
 \else \expandafter \@secondoftwo
 \fi
}%
\providecommand \@ifx [1]{%
 \ifx #1\expandafter \@firstoftwo
 \else \expandafter \@secondoftwo
 \fi
}%
\providecommand \natexlab [1]{#1}%
\providecommand \enquote  [1]{``#1''}%
\providecommand \bibnamefont  [1]{#1}%
\providecommand \bibfnamefont [1]{#1}%
\providecommand \citenamefont [1]{#1}%
\providecommand \href@noop [0]{\@secondoftwo}%
\providecommand \href [0]{\begingroup \@sanitize@url \@href}%
\providecommand \@href[1]{\@@startlink{#1}\@@href}%
\providecommand \@@href[1]{\endgroup#1\@@endlink}%
\providecommand \@sanitize@url [0]{\catcode `\\12\catcode `\$12\catcode
  `\&12\catcode `\#12\catcode `\^12\catcode `\_12\catcode `\%12\relax}%
\providecommand \@@startlink[1]{}%
\providecommand \@@endlink[0]{}%
\providecommand \url  [0]{\begingroup\@sanitize@url \@url }%
\providecommand \@url [1]{\endgroup\@href {#1}{\urlprefix }}%
\providecommand \urlprefix  [0]{URL }%
\providecommand \Eprint [0]{\href }%
\providecommand \doibase [0]{http://dx.doi.org/}%
\providecommand \selectlanguage [0]{\@gobble}%
\providecommand \bibinfo  [0]{\@secondoftwo}%
\providecommand \bibfield  [0]{\@secondoftwo}%
\providecommand \translation [1]{[#1]}%
\providecommand \BibitemOpen [0]{}%
\providecommand \bibitemStop [0]{}%
\providecommand \bibitemNoStop [0]{.\EOS\space}%
\providecommand \EOS [0]{\spacefactor3000\relax}%
\providecommand \BibitemShut  [1]{\csname bibitem#1\endcsname}%
\let\auto@bib@innerbib\@empty
\bibitem [{\citenamefont {Feynman}(1982)}]{feynman1982simulating}%
  \BibitemOpen
  \bibfield  {author} {\bibinfo {author} {\bibfnamefont {Richard~P.}\
  \bibnamefont {Feynman}},\ }\bibfield  {title} {\enquote {\bibinfo {title}
  {Simulating physics with computers},}\ }\href {\doibase 10.1007/BF02650179}
  {\bibfield  {journal} {\bibinfo  {journal} {International Journal of
  Theoretical Physics}\ }\textbf {\bibinfo {volume} {21}},\ \bibinfo {pages}
  {467--488} (\bibinfo {year} {1982})}\BibitemShut {NoStop}%
\bibitem [{\citenamefont {Lloyd}(1996)}]{lloyd1996universal}%
  \BibitemOpen
  \bibfield  {author} {\bibinfo {author} {\bibfnamefont {Seth}\ \bibnamefont
  {Lloyd}},\ }\bibfield  {title} {\enquote {\bibinfo {title} {Universal quantum
  simulators},}\ }\href {\doibase 10.1126/science.273.5278.1073} {\bibfield
  {journal} {\bibinfo  {journal} {Science}\ }\textbf {\bibinfo {volume}
  {273}},\ \bibinfo {pages} {1073--1078} (\bibinfo {year} {1996})}\BibitemShut
  {NoStop}%
\bibitem [{\citenamefont {Abrams}\ and\ \citenamefont
  {Lloyd}(1997)}]{abrams1997simulation}%
  \BibitemOpen
  \bibfield  {author} {\bibinfo {author} {\bibfnamefont {Daniel~S.}\
  \bibnamefont {Abrams}}\ and\ \bibinfo {author} {\bibfnamefont {Seth}\
  \bibnamefont {Lloyd}},\ }\bibfield  {title} {\enquote {\bibinfo {title}
  {Simulation of many-body fermi systems on a universal quantum computer},}\
  }\href {\doibase 10.1103/PhysRevLett.79.2586} {\bibfield  {journal} {\bibinfo
   {journal} {Phys. Rev. Lett.}\ }\textbf {\bibinfo {volume} {79}},\ \bibinfo
  {pages} {2586--2589} (\bibinfo {year} {1997})}\BibitemShut {NoStop}%
\bibitem [{\citenamefont {Aspuru-Guzik}\ \emph {et~al.}(2005)\citenamefont
  {Aspuru-Guzik}, \citenamefont {Dutoi}, \citenamefont {Love},\ and\
  \citenamefont {Head-Gordon}}]{aspuru-guzik2005simulated}%
  \BibitemOpen
  \bibfield  {author} {\bibinfo {author} {\bibfnamefont {Al{\'a}n}\
  \bibnamefont {Aspuru-Guzik}}, \bibinfo {author} {\bibfnamefont {Anthony~D.}\
  \bibnamefont {Dutoi}}, \bibinfo {author} {\bibfnamefont {Peter~J.}\
  \bibnamefont {Love}}, \ and\ \bibinfo {author} {\bibfnamefont {Martin}\
  \bibnamefont {Head-Gordon}},\ }\bibfield  {title} {\enquote {\bibinfo {title}
  {Simulated quantum computation of molecular energies},}\ }\href {\doibase
  10.1126/science.1113479} {\bibfield  {journal} {\bibinfo  {journal}
  {Science}\ }\textbf {\bibinfo {volume} {309}},\ \bibinfo {pages} {1704--1707}
  (\bibinfo {year} {2005})}\BibitemShut {NoStop}%
\bibitem [{\citenamefont {Georgescu}\ \emph {et~al.}(2014)\citenamefont
  {Georgescu}, \citenamefont {Ashhab},\ and\ \citenamefont
  {Nori}}]{georgescu2014quantum}%
  \BibitemOpen
  \bibfield  {author} {\bibinfo {author} {\bibfnamefont {I.~M.}\ \bibnamefont
  {Georgescu}}, \bibinfo {author} {\bibfnamefont {S.}~\bibnamefont {Ashhab}}, \
  and\ \bibinfo {author} {\bibfnamefont {Franco}\ \bibnamefont {Nori}},\
  }\bibfield  {title} {\enquote {\bibinfo {title} {Quantum simulation},}\
  }\href {\doibase 10.1103/RevModPhys.86.153} {\bibfield  {journal} {\bibinfo
  {journal} {Rev. Mod. Phys.}\ }\textbf {\bibinfo {volume} {86}},\ \bibinfo
  {pages} {153--185} (\bibinfo {year} {2014})}\BibitemShut {NoStop}%
\bibitem [{\citenamefont {Kitaev}(2003)}]{kitaev2003fault-tolerant}%
  \BibitemOpen
  \bibfield  {author} {\bibinfo {author} {\bibfnamefont {A.Yu.}\ \bibnamefont
  {Kitaev}},\ }\bibfield  {title} {\enquote {\bibinfo {title} {Fault-tolerant
  quantum computation by anyons},}\ }\href {\doibase
  https://doi.org/10.1016/S0003-4916(02)00018-0} {\bibfield  {journal}
  {\bibinfo  {journal} {Annals of Physics}\ }\textbf {\bibinfo {volume}
  {303}},\ \bibinfo {pages} {2--30} (\bibinfo {year} {2003})}\BibitemShut
  {NoStop}%
\bibitem [{\citenamefont {R\"opke}\ \emph {et~al.}(1998)\citenamefont
  {R\"opke}, \citenamefont {Schnell}, \citenamefont {Schuck},\ and\
  \citenamefont {Nozi\`eres}}]{ropke1998four-particle}%
  \BibitemOpen
  \bibfield  {author} {\bibinfo {author} {\bibfnamefont {G.}~\bibnamefont
  {R\"opke}}, \bibinfo {author} {\bibfnamefont {A.}~\bibnamefont {Schnell}},
  \bibinfo {author} {\bibfnamefont {P.}~\bibnamefont {Schuck}}, \ and\ \bibinfo
  {author} {\bibfnamefont {P.}~\bibnamefont {Nozi\`eres}},\ }\bibfield  {title}
  {\enquote {\bibinfo {title} {Four-particle condensate in strongly coupled
  fermion systems},}\ }\href {\doibase 10.1103/PhysRevLett.80.3177} {\bibfield
  {journal} {\bibinfo  {journal} {Phys. Rev. Lett.}\ }\textbf {\bibinfo
  {volume} {80}},\ \bibinfo {pages} {3177--3180} (\bibinfo {year}
  {1998})}\BibitemShut {NoStop}%
\bibitem [{\citenamefont {Funaki}\ \emph {et~al.}(2008)\citenamefont {Funaki},
  \citenamefont {Yamada}, \citenamefont {Horiuchi}, \citenamefont {R\"opke},
  \citenamefont {Schuck},\ and\ \citenamefont
  {Tohsaki}}]{funaki2008alpha-particle}%
  \BibitemOpen
  \bibfield  {author} {\bibinfo {author} {\bibfnamefont {Y.}~\bibnamefont
  {Funaki}}, \bibinfo {author} {\bibfnamefont {T.}~\bibnamefont {Yamada}},
  \bibinfo {author} {\bibfnamefont {H.}~\bibnamefont {Horiuchi}}, \bibinfo
  {author} {\bibfnamefont {G.}~\bibnamefont {R\"opke}}, \bibinfo {author}
  {\bibfnamefont {P.}~\bibnamefont {Schuck}}, \ and\ \bibinfo {author}
  {\bibfnamefont {A.}~\bibnamefont {Tohsaki}},\ }\bibfield  {title} {\enquote
  {\bibinfo {title} {$\ensuremath{\alpha}$-particle condensation in
  $^{16}\mathrm{O}$ studied with a full four-body orthogonality condition model
  calculation},}\ }\href {\doibase 10.1103/PhysRevLett.101.082502} {\bibfield
  {journal} {\bibinfo  {journal} {Phys. Rev. Lett.}\ }\textbf {\bibinfo
  {volume} {101}},\ \bibinfo {pages} {082502} (\bibinfo {year}
  {2008})}\BibitemShut {NoStop}%
\bibitem [{\citenamefont {Schuck}(2013)}]{schuck2013alpha-particle}%
  \BibitemOpen
  \bibfield  {author} {\bibinfo {author} {\bibfnamefont {P}~\bibnamefont
  {Schuck}},\ }\bibfield  {title} {\enquote {\bibinfo {title} {Alpha-particle
  condensation in nuclear systems: present status and perspectives},}\ }\href
  {\doibase 10.1088/1742-6596/436/1/012065} {\bibfield  {journal} {\bibinfo
  {journal} {Journal of Physics: Conference Series}\ }\textbf {\bibinfo
  {volume} {436}},\ \bibinfo {pages} {012065} (\bibinfo {year}
  {2013})}\BibitemShut {NoStop}%
\bibitem [{\citenamefont {Volovik}(2024)}]{volovik2024fermionic}%
  \BibitemOpen
  \bibfield  {author} {\bibinfo {author} {\bibfnamefont {G.~E.}\ \bibnamefont
  {Volovik}},\ }\bibfield  {title} {\enquote {\bibinfo {title} {Fermionic
  quartet and vestigial gravity},}\ }\href {\doibase 10.1134/S002136402460006X}
  {\bibfield  {journal} {\bibinfo  {journal} {JETP Letters}\ }\textbf {\bibinfo
  {volume} {119}},\ \bibinfo {pages} {330--334} (\bibinfo {year}
  {2024})}\BibitemShut {NoStop}%
\bibitem [{\citenamefont {Mizel}\ and\ \citenamefont
  {Lidar}(2004)}]{mizel2004three}%
  \BibitemOpen
  \bibfield  {author} {\bibinfo {author} {\bibfnamefont {Ari}\ \bibnamefont
  {Mizel}}\ and\ \bibinfo {author} {\bibfnamefont {Daniel~A.}\ \bibnamefont
  {Lidar}},\ }\bibfield  {title} {\enquote {\bibinfo {title} {Three- and
  four-body interactions in spin-based quantum computers},}\ }\href {\doibase
  10.1103/PhysRevLett.92.077903} {\bibfield  {journal} {\bibinfo  {journal}
  {Phys. Rev. Lett.}\ }\textbf {\bibinfo {volume} {92}},\ \bibinfo {pages}
  {077903} (\bibinfo {year} {2004})}\BibitemShut {NoStop}%
\bibitem [{\citenamefont {Peng}\ \emph {et~al.}(2009)\citenamefont {Peng},
  \citenamefont {Zhang}, \citenamefont {Du},\ and\ \citenamefont
  {Suter}}]{peng2009quantum}%
  \BibitemOpen
  \bibfield  {author} {\bibinfo {author} {\bibfnamefont {Xinhua}\ \bibnamefont
  {Peng}}, \bibinfo {author} {\bibfnamefont {Jingfu}\ \bibnamefont {Zhang}},
  \bibinfo {author} {\bibfnamefont {Jiangfeng}\ \bibnamefont {Du}}, \ and\
  \bibinfo {author} {\bibfnamefont {Dieter}\ \bibnamefont {Suter}},\ }\bibfield
   {title} {\enquote {\bibinfo {title} {Quantum simulation of a system with
  competing two- and three-body interactions},}\ }\href {\doibase
  10.1103/PhysRevLett.103.140501} {\bibfield  {journal} {\bibinfo  {journal}
  {Phys. Rev. Lett.}\ }\textbf {\bibinfo {volume} {103}},\ \bibinfo {pages}
  {140501} (\bibinfo {year} {2009})}\BibitemShut {NoStop}%
\bibitem [{\citenamefont {Dai}\ \emph {et~al.}(2017)\citenamefont {Dai},
  \citenamefont {Yang}, \citenamefont {Reingruber}, \citenamefont {Sun},
  \citenamefont {Xu}, \citenamefont {Chen}, \citenamefont {Yuan},\ and\
  \citenamefont {Pan}}]{dai2017four-body}%
  \BibitemOpen
  \bibfield  {author} {\bibinfo {author} {\bibfnamefont {Han-Ning}\
  \bibnamefont {Dai}}, \bibinfo {author} {\bibfnamefont {Bing}\ \bibnamefont
  {Yang}}, \bibinfo {author} {\bibfnamefont {Andreas}\ \bibnamefont
  {Reingruber}}, \bibinfo {author} {\bibfnamefont {Hui}\ \bibnamefont {Sun}},
  \bibinfo {author} {\bibfnamefont {Xiao-Fan}\ \bibnamefont {Xu}}, \bibinfo
  {author} {\bibfnamefont {Yu-Ao}\ \bibnamefont {Chen}}, \bibinfo {author}
  {\bibfnamefont {Zhen-Sheng}\ \bibnamefont {Yuan}}, \ and\ \bibinfo {author}
  {\bibfnamefont {Jian-Wei}\ \bibnamefont {Pan}},\ }\bibfield  {title}
  {\enquote {\bibinfo {title} {Four-body ring-exchange interactions and anyonic
  statistics within a minimal toric-code hamiltonian},}\ }\href {\doibase
  10.1038/nphys4243} {\bibfield  {journal} {\bibinfo  {journal} {Nature
  Physics}\ }\textbf {\bibinfo {volume} {13}},\ \bibinfo {pages} {1195--1200}
  (\bibinfo {year} {2017})}\BibitemShut {NoStop}%
\bibitem [{\citenamefont {Kumar}\ \emph {et~al.}(2020)\citenamefont {Kumar},
  \citenamefont {Zhang},\ and\ \citenamefont {Huang}}]{kumar2020large-scale}%
  \BibitemOpen
  \bibfield  {author} {\bibinfo {author} {\bibfnamefont {Santosh}\ \bibnamefont
  {Kumar}}, \bibinfo {author} {\bibfnamefont {He}~\bibnamefont {Zhang}}, \ and\
  \bibinfo {author} {\bibfnamefont {Yu-Ping}\ \bibnamefont {Huang}},\
  }\bibfield  {title} {\enquote {\bibinfo {title} {Large-scale ising emulation
  with four body interaction and all-to-all connections},}\ }\href {\doibase
  10.1038/s42005-020-0376-5} {\bibfield  {journal} {\bibinfo  {journal}
  {Communications Physics}\ }\textbf {\bibinfo {volume} {3}},\ \bibinfo {pages}
  {108} (\bibinfo {year} {2020})}\BibitemShut {NoStop}%
\bibitem [{\citenamefont {Zhang}\ \emph {et~al.}(2022)\citenamefont {Zhang},
  \citenamefont {Li}, \citenamefont {Zhang}, \citenamefont {Yuan},
  \citenamefont {Chen}, \citenamefont {Ren}, \citenamefont {Wang},
  \citenamefont {Song}, \citenamefont {Wang}, \citenamefont {Wang},
  \citenamefont {Zhu}, \citenamefont {Agarwal},\ and\ \citenamefont
  {Scully}}]{zhang2022synthesizing}%
  \BibitemOpen
  \bibfield  {author} {\bibinfo {author} {\bibfnamefont {Ke}~\bibnamefont
  {Zhang}}, \bibinfo {author} {\bibfnamefont {Hekang}\ \bibnamefont {Li}},
  \bibinfo {author} {\bibfnamefont {Pengfei}\ \bibnamefont {Zhang}}, \bibinfo
  {author} {\bibfnamefont {Jiale}\ \bibnamefont {Yuan}}, \bibinfo {author}
  {\bibfnamefont {Jinyan}\ \bibnamefont {Chen}}, \bibinfo {author}
  {\bibfnamefont {Wenhui}\ \bibnamefont {Ren}}, \bibinfo {author}
  {\bibfnamefont {Zhen}\ \bibnamefont {Wang}}, \bibinfo {author} {\bibfnamefont
  {Chao}\ \bibnamefont {Song}}, \bibinfo {author} {\bibfnamefont {Da-Wei}\
  \bibnamefont {Wang}}, \bibinfo {author} {\bibfnamefont {H.}~\bibnamefont
  {Wang}}, \bibinfo {author} {\bibfnamefont {Shiyao}\ \bibnamefont {Zhu}},
  \bibinfo {author} {\bibfnamefont {Girish~S.}\ \bibnamefont {Agarwal}}, \ and\
  \bibinfo {author} {\bibfnamefont {Marlan~O.}\ \bibnamefont {Scully}},\
  }\bibfield  {title} {\enquote {\bibinfo {title} {Synthesizing five-body
  interaction in a superconducting quantum circuit},}\ }\href {\doibase
  10.1103/PhysRevLett.128.190502} {\bibfield  {journal} {\bibinfo  {journal}
  {Phys. Rev. Lett.}\ }\textbf {\bibinfo {volume} {128}},\ \bibinfo {pages}
  {190502} (\bibinfo {year} {2022})}\BibitemShut {NoStop}%
\bibitem [{\citenamefont {Wu}(2005)}]{wu2005competing}%
  \BibitemOpen
  \bibfield  {author} {\bibinfo {author} {\bibfnamefont {Congjun}\ \bibnamefont
  {Wu}},\ }\bibfield  {title} {\enquote {\bibinfo {title} {Competing orders in
  one-dimensional spin-$3/2$ fermionic systems},}\ }\href {\doibase
  10.1103/PhysRevLett.95.266404} {\bibfield  {journal} {\bibinfo  {journal}
  {Phys. Rev. Lett.}\ }\textbf {\bibinfo {volume} {95}},\ \bibinfo {pages}
  {266404} (\bibinfo {year} {2005})}\BibitemShut {NoStop}%
\bibitem [{\citenamefont {Babaev}\ \emph {et~al.}(2004)\citenamefont {Babaev},
  \citenamefont {Sudb{\o}},\ and\ \citenamefont
  {Ashcroft}}]{babaev2004superconductor}%
  \BibitemOpen
  \bibfield  {author} {\bibinfo {author} {\bibfnamefont {Egor}\ \bibnamefont
  {Babaev}}, \bibinfo {author} {\bibfnamefont {Asle}\ \bibnamefont {Sudb{\o}}},
  \ and\ \bibinfo {author} {\bibfnamefont {N.~W.}\ \bibnamefont {Ashcroft}},\
  }\bibfield  {title} {\enquote {\bibinfo {title} {A superconductor to
  superfluid phase transition in liquid metallic hydrogen},}\ }\href {\doibase
  10.1038/nature02910} {\bibfield  {journal} {\bibinfo  {journal} {Nature}\
  }\textbf {\bibinfo {volume} {431}},\ \bibinfo {pages} {666--668} (\bibinfo
  {year} {2004})}\BibitemShut {NoStop}%
\bibitem [{\citenamefont {Berg}\ \emph {et~al.}(2009)\citenamefont {Berg},
  \citenamefont {Fradkin},\ and\ \citenamefont {Kivelson}}]{berg2009charge-4e}%
  \BibitemOpen
  \bibfield  {author} {\bibinfo {author} {\bibfnamefont {Erez}\ \bibnamefont
  {Berg}}, \bibinfo {author} {\bibfnamefont {Eduardo}\ \bibnamefont {Fradkin}},
  \ and\ \bibinfo {author} {\bibfnamefont {Steven~A.}\ \bibnamefont
  {Kivelson}},\ }\bibfield  {title} {\enquote {\bibinfo {title} {Charge-4e
  superconductivity from pair-density-wave order in certain high-temperature
  superconductors},}\ }\href {\doibase 10.1038/nphys1389} {\bibfield  {journal}
  {\bibinfo  {journal} {Nature Physics}\ }\textbf {\bibinfo {volume} {5}},\
  \bibinfo {pages} {830--833} (\bibinfo {year} {2009})}\BibitemShut {NoStop}%
\bibitem [{\citenamefont {Herland}\ \emph {et~al.}(2010)\citenamefont
  {Herland}, \citenamefont {Babaev},\ and\ \citenamefont
  {Sudb\o{}}}]{herland2010phase}%
  \BibitemOpen
  \bibfield  {author} {\bibinfo {author} {\bibfnamefont {Egil~V.}\ \bibnamefont
  {Herland}}, \bibinfo {author} {\bibfnamefont {Egor}\ \bibnamefont {Babaev}},
  \ and\ \bibinfo {author} {\bibfnamefont {Asle}\ \bibnamefont {Sudb\o{}}},\
  }\bibfield  {title} {\enquote {\bibinfo {title} {Phase transitions in a three
  dimensional $u(1)\ifmmode\times\else\texttimes\fi{}u(1)$ lattice london
  superconductor: Metallic superfluid and charge-$4e$ superconducting
  states},}\ }\href {\doibase 10.1103/PhysRevB.82.134511} {\bibfield  {journal}
  {\bibinfo  {journal} {Phys. Rev. B}\ }\textbf {\bibinfo {volume} {82}},\
  \bibinfo {pages} {134511} (\bibinfo {year} {2010})}\BibitemShut {NoStop}%
\bibitem [{\citenamefont {Lee}(2014)}]{lee2014amperean}%
  \BibitemOpen
  \bibfield  {author} {\bibinfo {author} {\bibfnamefont {Patrick~A.}\
  \bibnamefont {Lee}},\ }\bibfield  {title} {\enquote {\bibinfo {title}
  {Amperean pairing and the pseudogap phase of cuprate superconductors},}\
  }\href {\doibase 10.1103/PhysRevX.4.031017} {\bibfield  {journal} {\bibinfo
  {journal} {Phys. Rev. X}\ }\textbf {\bibinfo {volume} {4}},\ \bibinfo {pages}
  {031017} (\bibinfo {year} {2014})}\BibitemShut {NoStop}%
\bibitem [{\citenamefont {Fradkin}\ \emph {et~al.}(2015)\citenamefont
  {Fradkin}, \citenamefont {Kivelson},\ and\ \citenamefont
  {Tranquada}}]{fradkin2015colloquium}%
  \BibitemOpen
  \bibfield  {author} {\bibinfo {author} {\bibfnamefont {Eduardo}\ \bibnamefont
  {Fradkin}}, \bibinfo {author} {\bibfnamefont {Steven~A.}\ \bibnamefont
  {Kivelson}}, \ and\ \bibinfo {author} {\bibfnamefont {John~M.}\ \bibnamefont
  {Tranquada}},\ }\bibfield  {title} {\enquote {\bibinfo {title} {Colloquium:
  Theory of intertwined orders in high temperature superconductors},}\ }\href
  {\doibase 10.1103/RevModPhys.87.457} {\bibfield  {journal} {\bibinfo
  {journal} {Rev. Mod. Phys.}\ }\textbf {\bibinfo {volume} {87}},\ \bibinfo
  {pages} {457--482} (\bibinfo {year} {2015})}\BibitemShut {NoStop}%
\bibitem [{\citenamefont {Li}\ \emph {et~al.}(2007)\citenamefont {Li},
  \citenamefont {H\"ucker}, \citenamefont {Gu}, \citenamefont {Tsvelik},\ and\
  \citenamefont {Tranquada}}]{liPRL2007two-dimensional}%
  \BibitemOpen
  \bibfield  {author} {\bibinfo {author} {\bibfnamefont {Q.}~\bibnamefont
  {Li}}, \bibinfo {author} {\bibfnamefont {M.}~\bibnamefont {H\"ucker}},
  \bibinfo {author} {\bibfnamefont {G.~D.}\ \bibnamefont {Gu}}, \bibinfo
  {author} {\bibfnamefont {A.~M.}\ \bibnamefont {Tsvelik}}, \ and\ \bibinfo
  {author} {\bibfnamefont {J.~M.}\ \bibnamefont {Tranquada}},\ }\bibfield
  {title} {\enquote {\bibinfo {title} {Two-dimensional superconducting
  fluctuations in stripe-ordered
  ${\mathrm{la}}_{1.875}{\mathrm{ba}}_{0.125}{\mathrm{cuo}}_{4}$},}\ }\href
  {\doibase 10.1103/PhysRevLett.99.067001} {\bibfield  {journal} {\bibinfo
  {journal} {Phys. Rev. Lett.}\ }\textbf {\bibinfo {volume} {99}},\ \bibinfo
  {pages} {067001} (\bibinfo {year} {2007})}\BibitemShut {NoStop}%
\bibitem [{\citenamefont {Ding}\ \emph {et~al.}(2008)\citenamefont {Ding},
  \citenamefont {Xiang}, \citenamefont {Zhang}, \citenamefont {Liu},\ and\
  \citenamefont {Li}}]{dingPRB2008two-dimensional}%
  \BibitemOpen
  \bibfield  {author} {\bibinfo {author} {\bibfnamefont {J.~F.}\ \bibnamefont
  {Ding}}, \bibinfo {author} {\bibfnamefont {X.~Q.}\ \bibnamefont {Xiang}},
  \bibinfo {author} {\bibfnamefont {Y.~Q.}\ \bibnamefont {Zhang}}, \bibinfo
  {author} {\bibfnamefont {H.}~\bibnamefont {Liu}}, \ and\ \bibinfo {author}
  {\bibfnamefont {X.~G.}\ \bibnamefont {Li}},\ }\bibfield  {title} {\enquote
  {\bibinfo {title} {Two-dimensional superconductivity in stripe-ordered
  ${\text{la}}_{1.6\ensuremath{-}x}{\text{nd}}_{0.4}{\text{sr}}_{x}{\text{cuo}}_{4}$
  single crystals},}\ }\href {\doibase 10.1103/PhysRevB.77.214524} {\bibfield
  {journal} {\bibinfo  {journal} {Phys. Rev. B}\ }\textbf {\bibinfo {volume}
  {77}},\ \bibinfo {pages} {214524} (\bibinfo {year} {2008})}\BibitemShut
  {NoStop}%
\bibitem [{\citenamefont {Lin}\ and\ \citenamefont
  {Nandkishore}(2021)}]{lin2021complex}%
  \BibitemOpen
  \bibfield  {author} {\bibinfo {author} {\bibfnamefont {Yu-Ping}\ \bibnamefont
  {Lin}}\ and\ \bibinfo {author} {\bibfnamefont {Rahul~M.}\ \bibnamefont
  {Nandkishore}},\ }\bibfield  {title} {\enquote {\bibinfo {title} {Complex
  charge density waves at van hove singularity on hexagonal lattices:
  Haldane-model phase diagram and potential realization in the kagome metals
  $a{V}_{3}{\mathrm{sb}}_{5}$ ($a$=k, rb, cs)},}\ }\href {\doibase
  10.1103/PhysRevB.104.045122} {\bibfield  {journal} {\bibinfo  {journal}
  {Phys. Rev. B}\ }\textbf {\bibinfo {volume} {104}},\ \bibinfo {pages}
  {045122} (\bibinfo {year} {2021})}\BibitemShut {NoStop}%
\bibitem [{\citenamefont {Chen}\ \emph {et~al.}(2021)\citenamefont {Chen},
  \citenamefont {Yang}, \citenamefont {Hu}, \citenamefont {Zhao}, \citenamefont
  {Yuan}, \citenamefont {Xing}, \citenamefont {Qian}, \citenamefont {Huang},
  \citenamefont {Li}, \citenamefont {Ye}, \citenamefont {Ma}, \citenamefont
  {Ni}, \citenamefont {Zhang}, \citenamefont {Yin}, \citenamefont {Gong},
  \citenamefont {Tu}, \citenamefont {Lei}, \citenamefont {Tan}, \citenamefont
  {Zhou}, \citenamefont {Shen}, \citenamefont {Dong}, \citenamefont {Yan},
  \citenamefont {Wang},\ and\ \citenamefont {Gao}}]{chen2021roton}%
  \BibitemOpen
  \bibfield  {author} {\bibinfo {author} {\bibfnamefont {Hui}\ \bibnamefont
  {Chen}}, \bibinfo {author} {\bibfnamefont {Haitao}\ \bibnamefont {Yang}},
  \bibinfo {author} {\bibfnamefont {Bin}\ \bibnamefont {Hu}}, \bibinfo {author}
  {\bibfnamefont {Zhen}\ \bibnamefont {Zhao}}, \bibinfo {author} {\bibfnamefont
  {Jie}\ \bibnamefont {Yuan}}, \bibinfo {author} {\bibfnamefont {Yuqing}\
  \bibnamefont {Xing}}, \bibinfo {author} {\bibfnamefont {Guojian}\
  \bibnamefont {Qian}}, \bibinfo {author} {\bibfnamefont {Zihao}\ \bibnamefont
  {Huang}}, \bibinfo {author} {\bibfnamefont {Geng}\ \bibnamefont {Li}},
  \bibinfo {author} {\bibfnamefont {Yuhan}\ \bibnamefont {Ye}}, \bibinfo
  {author} {\bibfnamefont {Sheng}\ \bibnamefont {Ma}}, \bibinfo {author}
  {\bibfnamefont {Shunli}\ \bibnamefont {Ni}}, \bibinfo {author} {\bibfnamefont
  {Hua}\ \bibnamefont {Zhang}}, \bibinfo {author} {\bibfnamefont {Qiangwei}\
  \bibnamefont {Yin}}, \bibinfo {author} {\bibfnamefont {Chunsheng}\
  \bibnamefont {Gong}}, \bibinfo {author} {\bibfnamefont {Zhijun}\ \bibnamefont
  {Tu}}, \bibinfo {author} {\bibfnamefont {Hechang}\ \bibnamefont {Lei}},
  \bibinfo {author} {\bibfnamefont {Hengxin}\ \bibnamefont {Tan}}, \bibinfo
  {author} {\bibfnamefont {Sen}\ \bibnamefont {Zhou}}, \bibinfo {author}
  {\bibfnamefont {Chengmin}\ \bibnamefont {Shen}}, \bibinfo {author}
  {\bibfnamefont {Xiaoli}\ \bibnamefont {Dong}}, \bibinfo {author}
  {\bibfnamefont {Binghai}\ \bibnamefont {Yan}}, \bibinfo {author}
  {\bibfnamefont {Ziqiang}\ \bibnamefont {Wang}}, \ and\ \bibinfo {author}
  {\bibfnamefont {Hong-Jun}\ \bibnamefont {Gao}},\ }\bibfield  {title}
  {\enquote {\bibinfo {title} {Roton pair density wave in a strong-coupling
  kagome superconductor},}\ }\href {\doibase 10.1038/s41586-021-03983-5}
  {\bibfield  {journal} {\bibinfo  {journal} {Nature}\ }\textbf {\bibinfo
  {volume} {599}},\ \bibinfo {pages} {222--228} (\bibinfo {year}
  {2021})}\BibitemShut {NoStop}%
\bibitem [{\citenamefont {Zhou}\ and\ \citenamefont
  {Wang}(2022)}]{zhou2022chern}%
  \BibitemOpen
  \bibfield  {author} {\bibinfo {author} {\bibfnamefont {Sen}\ \bibnamefont
  {Zhou}}\ and\ \bibinfo {author} {\bibfnamefont {Ziqiang}\ \bibnamefont
  {Wang}},\ }\bibfield  {title} {\enquote {\bibinfo {title} {Chern fermi
  pocket, topological pair density wave, and charge-4e and charge-6e
  superconductivity in kagom{\'e}superconductors},}\ }\href {\doibase
  10.1038/s41467-022-34832-2} {\bibfield  {journal} {\bibinfo  {journal}
  {Nature Communications}\ }\textbf {\bibinfo {volume} {13}},\ \bibinfo {pages}
  {7288} (\bibinfo {year} {2022})}\BibitemShut {NoStop}%
\bibitem [{\citenamefont {Li}\ \emph {et~al.}(2023)\citenamefont {Li},
  \citenamefont {Oh}, \citenamefont {Kang}, \citenamefont {Zhao}, \citenamefont
  {Ortiz}, \citenamefont {Oey}, \citenamefont {Fang}, \citenamefont {Ren},
  \citenamefont {Jozwiak}, \citenamefont {Bostwick}, \citenamefont {Rotenberg},
  \citenamefont {Checkelsky}, \citenamefont {Wang}, \citenamefont {Wilson},
  \citenamefont {Comin},\ and\ \citenamefont {Zeljkovic}}]{liPRX2023small}%
  \BibitemOpen
  \bibfield  {author} {\bibinfo {author} {\bibfnamefont {Hong}\ \bibnamefont
  {Li}}, \bibinfo {author} {\bibfnamefont {Dongjin}\ \bibnamefont {Oh}},
  \bibinfo {author} {\bibfnamefont {Mingu}\ \bibnamefont {Kang}}, \bibinfo
  {author} {\bibfnamefont {He}~\bibnamefont {Zhao}}, \bibinfo {author}
  {\bibfnamefont {Brenden~R.}\ \bibnamefont {Ortiz}}, \bibinfo {author}
  {\bibfnamefont {Yuzki}\ \bibnamefont {Oey}}, \bibinfo {author} {\bibfnamefont
  {Shiang}\ \bibnamefont {Fang}}, \bibinfo {author} {\bibfnamefont {Zheng}\
  \bibnamefont {Ren}}, \bibinfo {author} {\bibfnamefont {Chris}\ \bibnamefont
  {Jozwiak}}, \bibinfo {author} {\bibfnamefont {Aaron}\ \bibnamefont
  {Bostwick}}, \bibinfo {author} {\bibfnamefont {Eli}\ \bibnamefont
  {Rotenberg}}, \bibinfo {author} {\bibfnamefont {Joseph~G.}\ \bibnamefont
  {Checkelsky}}, \bibinfo {author} {\bibfnamefont {Ziqiang}\ \bibnamefont
  {Wang}}, \bibinfo {author} {\bibfnamefont {Stephen~D.}\ \bibnamefont
  {Wilson}}, \bibinfo {author} {\bibfnamefont {Riccardo}\ \bibnamefont
  {Comin}}, \ and\ \bibinfo {author} {\bibfnamefont {Ilija}\ \bibnamefont
  {Zeljkovic}},\ }\bibfield  {title} {\enquote {\bibinfo {title} {Small fermi
  pockets intertwined with charge stripes and pair density wave order in a
  kagome superconductor},}\ }\href {\doibase 10.1103/PhysRevX.13.031030}
  {\bibfield  {journal} {\bibinfo  {journal} {Phys. Rev. X}\ }\textbf {\bibinfo
  {volume} {13}},\ \bibinfo {pages} {031030} (\bibinfo {year}
  {2023})}\BibitemShut {NoStop}%
\bibitem [{\citenamefont {Guguchia}\ \emph {et~al.}(2023)\citenamefont
  {Guguchia}, \citenamefont {Khasanov},\ and\ \citenamefont
  {Luetkens}}]{guguchia2023unconventional}%
  \BibitemOpen
  \bibfield  {author} {\bibinfo {author} {\bibfnamefont {Z.}~\bibnamefont
  {Guguchia}}, \bibinfo {author} {\bibfnamefont {R.}~\bibnamefont {Khasanov}},
  \ and\ \bibinfo {author} {\bibfnamefont {H.}~\bibnamefont {Luetkens}},\
  }\bibfield  {title} {\enquote {\bibinfo {title} {Unconventional charge order
  and superconductivity in kagome-lattice systems as seen by muon-spin
  rotation},}\ }\href {\doibase 10.1038/s41535-023-00574-7} {\bibfield
  {journal} {\bibinfo  {journal} {npj Quantum Materials}\ }\textbf {\bibinfo
  {volume} {8}},\ \bibinfo {pages} {41} (\bibinfo {year} {2023})}\BibitemShut
  {NoStop}%
\bibitem [{\citenamefont {Yu}(2023)}]{yu2023nondegenerate}%
  \BibitemOpen
  \bibfield  {author} {\bibinfo {author} {\bibfnamefont {Yue}\ \bibnamefont
  {Yu}},\ }\bibfield  {title} {\enquote {\bibinfo {title} {Nondegenerate
  surface pair density wave in the kagome superconductor
  ${\mathrm{csv}}_{3}{\mathrm{sb}}_{5}$: Application to vestigial orders},}\
  }\href {\doibase 10.1103/PhysRevB.108.054517} {\bibfield  {journal} {\bibinfo
   {journal} {Phys. Rev. B}\ }\textbf {\bibinfo {volume} {108}},\ \bibinfo
  {pages} {054517} (\bibinfo {year} {2023})}\BibitemShut {NoStop}%
\bibitem [{\citenamefont {Jian}\ \emph {et~al.}(2021)\citenamefont {Jian},
  \citenamefont {Huang},\ and\ \citenamefont {Yao}}]{jian2021charge-4e}%
  \BibitemOpen
  \bibfield  {author} {\bibinfo {author} {\bibfnamefont {Shao-Kai}\
  \bibnamefont {Jian}}, \bibinfo {author} {\bibfnamefont {Yingyi}\ \bibnamefont
  {Huang}}, \ and\ \bibinfo {author} {\bibfnamefont {Hong}\ \bibnamefont
  {Yao}},\ }\bibfield  {title} {\enquote {\bibinfo {title} {Charge-$4e$
  superconductivity from nematic superconductors in two and three
  dimensions},}\ }\href {\doibase 10.1103/PhysRevLett.127.227001} {\bibfield
  {journal} {\bibinfo  {journal} {Phys. Rev. Lett.}\ }\textbf {\bibinfo
  {volume} {127}},\ \bibinfo {pages} {227001} (\bibinfo {year}
  {2021})}\BibitemShut {NoStop}%
\bibitem [{\citenamefont {Fernandes}\ and\ \citenamefont
  {Fu}(2021)}]{fernandes2021charge-4e}%
  \BibitemOpen
  \bibfield  {author} {\bibinfo {author} {\bibfnamefont {Rafael~M.}\
  \bibnamefont {Fernandes}}\ and\ \bibinfo {author} {\bibfnamefont {Liang}\
  \bibnamefont {Fu}},\ }\bibfield  {title} {\enquote {\bibinfo {title}
  {Charge-$4e$ superconductivity from multicomponent nematic pairing:
  Application to twisted bilayer graphene},}\ }\href {\doibase
  10.1103/PhysRevLett.127.047001} {\bibfield  {journal} {\bibinfo  {journal}
  {Phys. Rev. Lett.}\ }\textbf {\bibinfo {volume} {127}},\ \bibinfo {pages}
  {047001} (\bibinfo {year} {2021})}\BibitemShut {NoStop}%
\bibitem [{\citenamefont {Liu}\ \emph {et~al.}(2023)\citenamefont {Liu},
  \citenamefont {Zhou}, \citenamefont {Wu},\ and\ \citenamefont
  {Yang}}]{liu2023charge-4e}%
  \BibitemOpen
  \bibfield  {author} {\bibinfo {author} {\bibfnamefont {Yu-Bo}\ \bibnamefont
  {Liu}}, \bibinfo {author} {\bibfnamefont {Jing}\ \bibnamefont {Zhou}},
  \bibinfo {author} {\bibfnamefont {Congjun}\ \bibnamefont {Wu}}, \ and\
  \bibinfo {author} {\bibfnamefont {Fan}\ \bibnamefont {Yang}},\ }\bibfield
  {title} {\enquote {\bibinfo {title} {Charge-4e superconductivity and chiral
  metal in 45$\,^{\circ}$-twisted bilayer cuprates and related bilayers},}\
  }\href {\doibase 10.1038/s41467-023-43782-2} {\bibfield  {journal} {\bibinfo
  {journal} {Nature Communications}\ }\textbf {\bibinfo {volume} {14}},\
  \bibinfo {pages} {7926} (\bibinfo {year} {2023})}\BibitemShut {NoStop}%
\bibitem [{\citenamefont {Fernandes}\ \emph {et~al.}(2014)\citenamefont
  {Fernandes}, \citenamefont {Chubukov},\ and\ \citenamefont
  {Schmalian}}]{fernandes2014what}%
  \BibitemOpen
  \bibfield  {author} {\bibinfo {author} {\bibfnamefont {R.~M.}\ \bibnamefont
  {Fernandes}}, \bibinfo {author} {\bibfnamefont {A.~V.}\ \bibnamefont
  {Chubukov}}, \ and\ \bibinfo {author} {\bibfnamefont {J.}~\bibnamefont
  {Schmalian}},\ }\bibfield  {title} {\enquote {\bibinfo {title} {What drives
  nematic order in iron-based superconductors?}}\ }\href {\doibase
  10.1038/nphys2877} {\bibfield  {journal} {\bibinfo  {journal} {Nature
  Physics}\ }\textbf {\bibinfo {volume} {10}},\ \bibinfo {pages} {97--104}
  (\bibinfo {year} {2014})}\BibitemShut {NoStop}%
\bibitem [{\citenamefont {Matano}\ \emph {et~al.}(2016)\citenamefont {Matano},
  \citenamefont {Kriener}, \citenamefont {Segawa}, \citenamefont {Ando},\ and\
  \citenamefont {Zheng}}]{matano2016spin-rotation}%
  \BibitemOpen
  \bibfield  {author} {\bibinfo {author} {\bibfnamefont {K.}~\bibnamefont
  {Matano}}, \bibinfo {author} {\bibfnamefont {M.}~\bibnamefont {Kriener}},
  \bibinfo {author} {\bibfnamefont {K.}~\bibnamefont {Segawa}}, \bibinfo
  {author} {\bibfnamefont {Y.}~\bibnamefont {Ando}}, \ and\ \bibinfo {author}
  {\bibfnamefont {Guo-qing}\ \bibnamefont {Zheng}},\ }\bibfield  {title}
  {\enquote {\bibinfo {title} {Spin-rotation symmetry breaking in the
  superconducting state of cuxbi2se3},}\ }\href {\doibase 10.1038/nphys3781}
  {\bibfield  {journal} {\bibinfo  {journal} {Nature Physics}\ }\textbf
  {\bibinfo {volume} {12}},\ \bibinfo {pages} {852--854} (\bibinfo {year}
  {2016})}\BibitemShut {NoStop}%
\bibitem [{\citenamefont {Shen}\ \emph {et~al.}(2017)\citenamefont {Shen},
  \citenamefont {He}, \citenamefont {Yuan}, \citenamefont {Huang},
  \citenamefont {Cho}, \citenamefont {Lee}, \citenamefont {Hor}, \citenamefont
  {Law},\ and\ \citenamefont {Lortz}}]{shen2017nematic}%
  \BibitemOpen
  \bibfield  {author} {\bibinfo {author} {\bibfnamefont {Junying}\ \bibnamefont
  {Shen}}, \bibinfo {author} {\bibfnamefont {Wen-Yu}\ \bibnamefont {He}},
  \bibinfo {author} {\bibfnamefont {Noah Fan~Qi}\ \bibnamefont {Yuan}},
  \bibinfo {author} {\bibfnamefont {Zengle}\ \bibnamefont {Huang}}, \bibinfo
  {author} {\bibfnamefont {Chang-woo}\ \bibnamefont {Cho}}, \bibinfo {author}
  {\bibfnamefont {Seng~Huat}\ \bibnamefont {Lee}}, \bibinfo {author}
  {\bibfnamefont {Yew~San}\ \bibnamefont {Hor}}, \bibinfo {author}
  {\bibfnamefont {Kam~Tuen}\ \bibnamefont {Law}}, \ and\ \bibinfo {author}
  {\bibfnamefont {Rolf}\ \bibnamefont {Lortz}},\ }\bibfield  {title} {\enquote
  {\bibinfo {title} {Nematic topological superconducting phase in nb-doped
  bi2se3},}\ }\href {\doibase 10.1038/s41535-017-0064-1} {\bibfield  {journal}
  {\bibinfo  {journal} {npj Quantum Materials}\ }\textbf {\bibinfo {volume}
  {2}},\ \bibinfo {pages} {59} (\bibinfo {year} {2017})}\BibitemShut {NoStop}%
\bibitem [{\citenamefont {Grinenko}\ \emph {et~al.}(2021)\citenamefont
  {Grinenko}, \citenamefont {Weston}, \citenamefont {Caglieris}, \citenamefont
  {Wuttke}, \citenamefont {Hess}, \citenamefont {Gottschall}, \citenamefont
  {Maccari}, \citenamefont {Gorbunov}, \citenamefont {Zherlitsyn},
  \citenamefont {Wosnitza}, \citenamefont {Rydh}, \citenamefont {Kihou},
  \citenamefont {Lee}, \citenamefont {Sarkar}, \citenamefont {Dengre},
  \citenamefont {Garaud}, \citenamefont {Charnukha}, \citenamefont {H{\"u}hne},
  \citenamefont {Nielsch}, \citenamefont {B{\"u}chner}, \citenamefont
  {Klauss},\ and\ \citenamefont {Babaev}}]{grinenko2021state}%
  \BibitemOpen
  \bibfield  {author} {\bibinfo {author} {\bibfnamefont {Vadim}\ \bibnamefont
  {Grinenko}}, \bibinfo {author} {\bibfnamefont {Daniel}\ \bibnamefont
  {Weston}}, \bibinfo {author} {\bibfnamefont {Federico}\ \bibnamefont
  {Caglieris}}, \bibinfo {author} {\bibfnamefont {Christoph}\ \bibnamefont
  {Wuttke}}, \bibinfo {author} {\bibfnamefont {Christian}\ \bibnamefont
  {Hess}}, \bibinfo {author} {\bibfnamefont {Tino}\ \bibnamefont {Gottschall}},
  \bibinfo {author} {\bibfnamefont {Ilaria}\ \bibnamefont {Maccari}}, \bibinfo
  {author} {\bibfnamefont {Denis}\ \bibnamefont {Gorbunov}}, \bibinfo {author}
  {\bibfnamefont {Sergei}\ \bibnamefont {Zherlitsyn}}, \bibinfo {author}
  {\bibfnamefont {Jochen}\ \bibnamefont {Wosnitza}}, \bibinfo {author}
  {\bibfnamefont {Andreas}\ \bibnamefont {Rydh}}, \bibinfo {author}
  {\bibfnamefont {Kunihiro}\ \bibnamefont {Kihou}}, \bibinfo {author}
  {\bibfnamefont {Chul-Ho}\ \bibnamefont {Lee}}, \bibinfo {author}
  {\bibfnamefont {Rajib}\ \bibnamefont {Sarkar}}, \bibinfo {author}
  {\bibfnamefont {Shanu}\ \bibnamefont {Dengre}}, \bibinfo {author}
  {\bibfnamefont {Julien}\ \bibnamefont {Garaud}}, \bibinfo {author}
  {\bibfnamefont {Aliaksei}\ \bibnamefont {Charnukha}}, \bibinfo {author}
  {\bibfnamefont {Ruben}\ \bibnamefont {H{\"u}hne}}, \bibinfo {author}
  {\bibfnamefont {Kornelius}\ \bibnamefont {Nielsch}}, \bibinfo {author}
  {\bibfnamefont {Bernd}\ \bibnamefont {B{\"u}chner}}, \bibinfo {author}
  {\bibfnamefont {Hans-Henning}\ \bibnamefont {Klauss}}, \ and\ \bibinfo
  {author} {\bibfnamefont {Egor}\ \bibnamefont {Babaev}},\ }\bibfield  {title}
  {\enquote {\bibinfo {title} {State with spontaneously broken time-reversal
  symmetry above the superconducting phase transition},}\ }\href {\doibase
  10.1038/s41567-021-01350-9} {\bibfield  {journal} {\bibinfo  {journal}
  {Nature Physics}\ }\textbf {\bibinfo {volume} {17}},\ \bibinfo {pages}
  {1254--1259} (\bibinfo {year} {2021})}\BibitemShut {NoStop}%
\bibitem [{\citenamefont {Babaev}(2024)}]{babaev2024topological}%
  \BibitemOpen
  \bibfield  {author} {\bibinfo {author} {\bibfnamefont {Egor}\ \bibnamefont
  {Babaev}},\ }\bibfield  {title} {\enquote {\bibinfo {title} {Topological
  order in higher composites},}\ }\href {\doibase
  10.1103/PhysRevResearch.6.L032034} {\bibfield  {journal} {\bibinfo  {journal}
  {Phys. Rev. Res.}\ }\textbf {\bibinfo {volume} {6}},\ \bibinfo {pages}
  {L032034} (\bibinfo {year} {2024})}\BibitemShut {NoStop}%
\bibitem [{\citenamefont {Maccari}\ and\ \citenamefont
  {Babaev}(2022)}]{maccari2022effects}%
  \BibitemOpen
  \bibfield  {author} {\bibinfo {author} {\bibfnamefont {I.}~\bibnamefont
  {Maccari}}\ and\ \bibinfo {author} {\bibfnamefont {E.}~\bibnamefont
  {Babaev}},\ }\bibfield  {title} {\enquote {\bibinfo {title} {Effects of
  intercomponent couplings on the appearance of time-reversal symmetry breaking
  fermion-quadrupling states in two-component london models},}\ }\href
  {\doibase 10.1103/PhysRevB.105.214520} {\bibfield  {journal} {\bibinfo
  {journal} {Phys. Rev. B}\ }\textbf {\bibinfo {volume} {105}},\ \bibinfo
  {pages} {214520} (\bibinfo {year} {2022})}\BibitemShut {NoStop}%
\bibitem [{\citenamefont {Maccari}\ \emph {et~al.}(2023)\citenamefont
  {Maccari}, \citenamefont {Carlstr\"om},\ and\ \citenamefont
  {Babaev}}]{maccari2023prediction}%
  \BibitemOpen
  \bibfield  {author} {\bibinfo {author} {\bibfnamefont {I.}~\bibnamefont
  {Maccari}}, \bibinfo {author} {\bibfnamefont {J.}~\bibnamefont
  {Carlstr\"om}}, \ and\ \bibinfo {author} {\bibfnamefont {E.}~\bibnamefont
  {Babaev}},\ }\bibfield  {title} {\enquote {\bibinfo {title} {Prediction of
  time-reversal-symmetry breaking fermionic quadrupling condensate in twisted
  bilayer graphene},}\ }\href {\doibase 10.1103/PhysRevB.107.064501} {\bibfield
   {journal} {\bibinfo  {journal} {Phys. Rev. B}\ }\textbf {\bibinfo {volume}
  {107}},\ \bibinfo {pages} {064501} (\bibinfo {year} {2023})}\BibitemShut
  {NoStop}%
\bibitem [{\citenamefont {Dou\ifmmode~\mbox{\c{c}}\else \c{c}\fi{}ot}\ and\
  \citenamefont {Vidal}(2002)}]{doucot2002pairing}%
  \BibitemOpen
  \bibfield  {author} {\bibinfo {author} {\bibfnamefont {Benoit}\ \bibnamefont
  {Dou\ifmmode~\mbox{\c{c}}\else \c{c}\fi{}ot}}\ and\ \bibinfo {author}
  {\bibfnamefont {Julien}\ \bibnamefont {Vidal}},\ }\bibfield  {title}
  {\enquote {\bibinfo {title} {Pairing of cooper pairs in a fully frustrated
  josephson-junction chain},}\ }\href {\doibase 10.1103/PhysRevLett.88.227005}
  {\bibfield  {journal} {\bibinfo  {journal} {Phys. Rev. Lett.}\ }\textbf
  {\bibinfo {volume} {88}},\ \bibinfo {pages} {227005} (\bibinfo {year}
  {2002})}\BibitemShut {NoStop}%
\bibitem [{\citenamefont {Dou\ifmmode~\mbox{\c{c}}\else \c{c}\fi{}ot}\ \emph
  {et~al.}(2003)\citenamefont {Dou\ifmmode~\mbox{\c{c}}\else \c{c}\fi{}ot},
  \citenamefont {Feigel'man},\ and\ \citenamefont
  {Ioffe}}]{doucot2003topological}%
  \BibitemOpen
  \bibfield  {author} {\bibinfo {author} {\bibfnamefont {B.}~\bibnamefont
  {Dou\ifmmode~\mbox{\c{c}}\else \c{c}\fi{}ot}}, \bibinfo {author}
  {\bibfnamefont {M.~V.}\ \bibnamefont {Feigel'man}}, \ and\ \bibinfo {author}
  {\bibfnamefont {L.~B.}\ \bibnamefont {Ioffe}},\ }\bibfield  {title} {\enquote
  {\bibinfo {title} {Topological order in the insulating josephson junction
  array},}\ }\href {\doibase 10.1103/PhysRevLett.90.107003} {\bibfield
  {journal} {\bibinfo  {journal} {Phys. Rev. Lett.}\ }\textbf {\bibinfo
  {volume} {90}},\ \bibinfo {pages} {107003} (\bibinfo {year}
  {2003})}\BibitemShut {NoStop}%
\bibitem [{\citenamefont {Protopopov}\ and\ \citenamefont
  {Feigel'man}(2004)}]{protopopov2004anomalous}%
  \BibitemOpen
  \bibfield  {author} {\bibinfo {author} {\bibfnamefont {Ivan~V.}\ \bibnamefont
  {Protopopov}}\ and\ \bibinfo {author} {\bibfnamefont {Mikhail~V.}\
  \bibnamefont {Feigel'man}},\ }\bibfield  {title} {\enquote {\bibinfo {title}
  {Anomalous periodicity of supercurrent in long frustrated josephson-junction
  rhombi chains},}\ }\href {\doibase 10.1103/PhysRevB.70.184519} {\bibfield
  {journal} {\bibinfo  {journal} {Phys. Rev. B}\ }\textbf {\bibinfo {volume}
  {70}},\ \bibinfo {pages} {184519} (\bibinfo {year} {2004})}\BibitemShut
  {NoStop}%
\bibitem [{\citenamefont {Dou\ifmmode~\mbox{\c{c}}\else \c{c}\fi{}ot}\ \emph
  {et~al.}(2005)\citenamefont {Dou\ifmmode~\mbox{\c{c}}\else \c{c}\fi{}ot},
  \citenamefont {Feigel'man}, \citenamefont {Ioffe},\ and\ \citenamefont
  {Ioselevich}}]{doucot2005protected}%
  \BibitemOpen
  \bibfield  {author} {\bibinfo {author} {\bibfnamefont {B.}~\bibnamefont
  {Dou\ifmmode~\mbox{\c{c}}\else \c{c}\fi{}ot}}, \bibinfo {author}
  {\bibfnamefont {M.~V.}\ \bibnamefont {Feigel'man}}, \bibinfo {author}
  {\bibfnamefont {L.~B.}\ \bibnamefont {Ioffe}}, \ and\ \bibinfo {author}
  {\bibfnamefont {A.~S.}\ \bibnamefont {Ioselevich}},\ }\bibfield  {title}
  {\enquote {\bibinfo {title} {Protected qubits and chern-simons theories in
  josephson junction arrays},}\ }\href {\doibase 10.1103/PhysRevB.71.024505}
  {\bibfield  {journal} {\bibinfo  {journal} {Phys. Rev. B}\ }\textbf {\bibinfo
  {volume} {71}},\ \bibinfo {pages} {024505} (\bibinfo {year}
  {2005})}\BibitemShut {NoStop}%
\bibitem [{\citenamefont {Rizzi}\ \emph {et~al.}(2006)\citenamefont {Rizzi},
  \citenamefont {Cataudella},\ and\ \citenamefont
  {Fazio}}]{rizzi20064e-condensation}%
  \BibitemOpen
  \bibfield  {author} {\bibinfo {author} {\bibfnamefont {Matteo}\ \bibnamefont
  {Rizzi}}, \bibinfo {author} {\bibfnamefont {Vittorio}\ \bibnamefont
  {Cataudella}}, \ and\ \bibinfo {author} {\bibfnamefont {Rosario}\
  \bibnamefont {Fazio}},\ }\bibfield  {title} {\enquote {\bibinfo {title}
  {$4e$-condensation in a fully frustrated josephson junction diamond chain},}\
  }\href {\doibase 10.1103/PhysRevB.73.100502} {\bibfield  {journal} {\bibinfo
  {journal} {Phys. Rev. B}\ }\textbf {\bibinfo {volume} {73}},\ \bibinfo
  {pages} {100502} (\bibinfo {year} {2006})}\BibitemShut {NoStop}%
\bibitem [{\citenamefont {Ciaccia}\ \emph {et~al.}(2024)\citenamefont
  {Ciaccia}, \citenamefont {Haller}, \citenamefont {Drachmann}, \citenamefont
  {Lindemann}, \citenamefont {Manfra}, \citenamefont {Schrade},\ and\
  \citenamefont {Sch{\"o}nenberger}}]{ciaccia2024charge-4e}%
  \BibitemOpen
  \bibfield  {author} {\bibinfo {author} {\bibfnamefont {Carlo}\ \bibnamefont
  {Ciaccia}}, \bibinfo {author} {\bibfnamefont {Roy}\ \bibnamefont {Haller}},
  \bibinfo {author} {\bibfnamefont {Asbj{\o}rn C.~C.}\ \bibnamefont
  {Drachmann}}, \bibinfo {author} {\bibfnamefont {Tyler}\ \bibnamefont
  {Lindemann}}, \bibinfo {author} {\bibfnamefont {Michael~J.}\ \bibnamefont
  {Manfra}}, \bibinfo {author} {\bibfnamefont {Constantin}\ \bibnamefont
  {Schrade}}, \ and\ \bibinfo {author} {\bibfnamefont {Christian}\ \bibnamefont
  {Sch{\"o}nenberger}},\ }\bibfield  {title} {\enquote {\bibinfo {title}
  {Charge-4e supercurrent in a two-dimensional inas-al
  superconductor-semiconductor heterostructure},}\ }\href {\doibase
  10.1038/s42005-024-01531-x} {\bibfield  {journal} {\bibinfo  {journal}
  {Communications Physics}\ }\textbf {\bibinfo {volume} {7}},\ \bibinfo {pages}
  {41} (\bibinfo {year} {2024})}\BibitemShut {NoStop}%
\bibitem [{\citenamefont {Cuevas}\ and\ \citenamefont
  {Pothier}(2007)}]{cuevas2007voltage-induced}%
  \BibitemOpen
  \bibfield  {author} {\bibinfo {author} {\bibfnamefont {J.~C.}\ \bibnamefont
  {Cuevas}}\ and\ \bibinfo {author} {\bibfnamefont {H.}~\bibnamefont
  {Pothier}},\ }\bibfield  {title} {\enquote {\bibinfo {title} {Voltage-induced
  shapiro steps in a superconducting multiterminal structure},}\ }\href
  {\doibase 10.1103/PhysRevB.75.174513} {\bibfield  {journal} {\bibinfo
  {journal} {Phys. Rev. B}\ }\textbf {\bibinfo {volume} {75}},\ \bibinfo
  {pages} {174513} (\bibinfo {year} {2007})}\BibitemShut {NoStop}%
\bibitem [{\citenamefont {Freyn}\ \emph {et~al.}(2011)\citenamefont {Freyn},
  \citenamefont {Dou\ifmmode~\mbox{\c{c}}\else \c{c}\fi{}ot}, \citenamefont
  {Feinberg},\ and\ \citenamefont {M\'elin}}]{freyn2011production}%
  \BibitemOpen
  \bibfield  {author} {\bibinfo {author} {\bibfnamefont {Axel}\ \bibnamefont
  {Freyn}}, \bibinfo {author} {\bibfnamefont {Benoit}\ \bibnamefont
  {Dou\ifmmode~\mbox{\c{c}}\else \c{c}\fi{}ot}}, \bibinfo {author}
  {\bibfnamefont {Denis}\ \bibnamefont {Feinberg}}, \ and\ \bibinfo {author}
  {\bibfnamefont {R\'egis}\ \bibnamefont {M\'elin}},\ }\bibfield  {title}
  {\enquote {\bibinfo {title} {Production of nonlocal quartets and
  phase-sensitive entanglement in a superconducting beam splitter},}\ }\href
  {\doibase 10.1103/PhysRevLett.106.257005} {\bibfield  {journal} {\bibinfo
  {journal} {Phys. Rev. Lett.}\ }\textbf {\bibinfo {volume} {106}},\ \bibinfo
  {pages} {257005} (\bibinfo {year} {2011})}\BibitemShut {NoStop}%
\bibitem [{\citenamefont {Jonckheere}\ \emph {et~al.}(2013)\citenamefont
  {Jonckheere}, \citenamefont {Rech}, \citenamefont {Martin}, \citenamefont
  {Dou\ifmmode~\mbox{\c{c}}\else \c{c}\fi{}ot}, \citenamefont {Feinberg},\ and\
  \citenamefont {M\'elin}}]{jonckheere2013multipair}%
  \BibitemOpen
  \bibfield  {author} {\bibinfo {author} {\bibfnamefont {T.}~\bibnamefont
  {Jonckheere}}, \bibinfo {author} {\bibfnamefont {J.}~\bibnamefont {Rech}},
  \bibinfo {author} {\bibfnamefont {T.}~\bibnamefont {Martin}}, \bibinfo
  {author} {\bibfnamefont {B.}~\bibnamefont {Dou\ifmmode~\mbox{\c{c}}\else
  \c{c}\fi{}ot}}, \bibinfo {author} {\bibfnamefont {D.}~\bibnamefont
  {Feinberg}}, \ and\ \bibinfo {author} {\bibfnamefont {R.}~\bibnamefont
  {M\'elin}},\ }\bibfield  {title} {\enquote {\bibinfo {title} {Multipair dc
  josephson resonances in a biased all-superconducting bijunction},}\ }\href
  {\doibase 10.1103/PhysRevB.87.214501} {\bibfield  {journal} {\bibinfo
  {journal} {Phys. Rev. B}\ }\textbf {\bibinfo {volume} {87}},\ \bibinfo
  {pages} {214501} (\bibinfo {year} {2013})}\BibitemShut {NoStop}%
\bibitem [{\citenamefont {Feinberg}\ \emph {et~al.}(2015)\citenamefont
  {Feinberg}, \citenamefont {Jonckheere}, \citenamefont {Rech}, \citenamefont
  {Martin}, \citenamefont {Dou{\c c}ot},\ and\ \citenamefont
  {M{\'e}lin}}]{feinberg2015quartets}%
  \BibitemOpen
  \bibfield  {author} {\bibinfo {author} {\bibfnamefont {Denis}\ \bibnamefont
  {Feinberg}}, \bibinfo {author} {\bibfnamefont {Thibaut}\ \bibnamefont
  {Jonckheere}}, \bibinfo {author} {\bibfnamefont {J{\'e}r{\^o}me}\
  \bibnamefont {Rech}}, \bibinfo {author} {\bibfnamefont {Thierry}\
  \bibnamefont {Martin}}, \bibinfo {author} {\bibfnamefont {Beno{\^\i}t}\
  \bibnamefont {Dou{\c c}ot}}, \ and\ \bibinfo {author} {\bibfnamefont
  {R{\'e}gis}\ \bibnamefont {M{\'e}lin}},\ }\bibfield  {title} {\enquote
  {\bibinfo {title} {Quartets and the current-phase structure of a double
  quantum dot superconducting bijunction at equilibrium},}\ }\href {\doibase
  10.1140/epjb/e2015-50849-3} {\bibfield  {journal} {\bibinfo  {journal} {The
  European Physical Journal B}\ }\textbf {\bibinfo {volume} {88}},\ \bibinfo
  {pages} {99} (\bibinfo {year} {2015})}\BibitemShut {NoStop}%
\bibitem [{\citenamefont {Cohen}\ \emph {et~al.}(2018)\citenamefont {Cohen},
  \citenamefont {Ronen}, \citenamefont {Kang}, \citenamefont {Heiblum},
  \citenamefont {Feinberg}, \citenamefont {M{\'e}lin},\ and\ \citenamefont
  {Shtrikman}}]{yonatan2018nonlocal}%
  \BibitemOpen
  \bibfield  {author} {\bibinfo {author} {\bibfnamefont {Yonatan}\ \bibnamefont
  {Cohen}}, \bibinfo {author} {\bibfnamefont {Yuval}\ \bibnamefont {Ronen}},
  \bibinfo {author} {\bibfnamefont {Jung-Hyun}\ \bibnamefont {Kang}}, \bibinfo
  {author} {\bibfnamefont {Moty}\ \bibnamefont {Heiblum}}, \bibinfo {author}
  {\bibfnamefont {Denis}\ \bibnamefont {Feinberg}}, \bibinfo {author}
  {\bibfnamefont {R{\'e}gis}\ \bibnamefont {M{\'e}lin}}, \ and\ \bibinfo
  {author} {\bibfnamefont {Hadas}\ \bibnamefont {Shtrikman}},\ }\bibfield
  {title} {\enquote {\bibinfo {title} {Nonlocal supercurrent of quartets in a
  three-terminal josephson junction},}\ }\href {\doibase
  10.1073/pnas.1800044115} {\bibfield  {journal} {\bibinfo  {journal}
  {Proceedings of the National Academy of Sciences}\ }\textbf {\bibinfo
  {volume} {115}},\ \bibinfo {pages} {6991--6994} (\bibinfo {year} {2018})},\
  \Eprint
  {http://arxiv.org/abs/https://www.pnas.org/doi/pdf/10.1073/pnas.1800044115}
  {https://www.pnas.org/doi/pdf/10.1073/pnas.1800044115} \BibitemShut {NoStop}%
\bibitem [{\citenamefont {Huang}\ \emph {et~al.}(2022)\citenamefont {Huang},
  \citenamefont {Ronen}, \citenamefont {M{\'e}lin}, \citenamefont {Feinberg},
  \citenamefont {Watanabe}, \citenamefont {Taniguchi},\ and\ \citenamefont
  {Kim}}]{huang2022evidence}%
  \BibitemOpen
  \bibfield  {author} {\bibinfo {author} {\bibfnamefont {Ko-Fan}\ \bibnamefont
  {Huang}}, \bibinfo {author} {\bibfnamefont {Yuval}\ \bibnamefont {Ronen}},
  \bibinfo {author} {\bibfnamefont {R{\'e}gis}\ \bibnamefont {M{\'e}lin}},
  \bibinfo {author} {\bibfnamefont {Denis}\ \bibnamefont {Feinberg}}, \bibinfo
  {author} {\bibfnamefont {Kenji}\ \bibnamefont {Watanabe}}, \bibinfo {author}
  {\bibfnamefont {Takashi}\ \bibnamefont {Taniguchi}}, \ and\ \bibinfo {author}
  {\bibfnamefont {Philip}\ \bibnamefont {Kim}},\ }\bibfield  {title} {\enquote
  {\bibinfo {title} {Evidence for 4e charge of cooper quartets in a biased
  multi-terminal graphene-based josephson junction},}\ }\href {\doibase
  10.1038/s41467-022-30732-7} {\bibfield  {journal} {\bibinfo  {journal}
  {Nature Communications}\ }\textbf {\bibinfo {volume} {13}},\ \bibinfo {pages}
  {3032} (\bibinfo {year} {2022})}\BibitemShut {NoStop}%
\bibitem [{\citenamefont {Melo}\ \emph {et~al.}(2022)\citenamefont {Melo},
  \citenamefont {Fatemi},\ and\ \citenamefont {Akhmerov}}]{melo2022multiplet}%
  \BibitemOpen
  \bibfield  {author} {\bibinfo {author} {\bibfnamefont {Andr{\'e}}\
  \bibnamefont {Melo}}, \bibinfo {author} {\bibfnamefont {Valla}\ \bibnamefont
  {Fatemi}}, \ and\ \bibinfo {author} {\bibfnamefont {Anton~R.}\ \bibnamefont
  {Akhmerov}},\ }\bibfield  {title} {\enquote {\bibinfo {title} {{Multiplet
  supercurrent in Josephson tunneling circuits}},}\ }\href {\doibase
  10.21468/SciPostPhys.12.1.017} {\bibfield  {journal} {\bibinfo  {journal}
  {SciPost Phys.}\ }\textbf {\bibinfo {volume} {12}},\ \bibinfo {pages} {017}
  (\bibinfo {year} {2022})}\BibitemShut {NoStop}%
\bibitem [{\citenamefont {Hamo}\ \emph {et~al.}(2016)\citenamefont {Hamo},
  \citenamefont {Benyamini}, \citenamefont {Shapir}, \citenamefont {Khivrich},
  \citenamefont {Waissman}, \citenamefont {Kaasbjerg}, \citenamefont {Oreg},
  \citenamefont {von Oppen},\ and\ \citenamefont {Ilani}}]{hamo2016electron}%
  \BibitemOpen
  \bibfield  {author} {\bibinfo {author} {\bibfnamefont {A.}~\bibnamefont
  {Hamo}}, \bibinfo {author} {\bibfnamefont {A.}~\bibnamefont {Benyamini}},
  \bibinfo {author} {\bibfnamefont {I.}~\bibnamefont {Shapir}}, \bibinfo
  {author} {\bibfnamefont {I.}~\bibnamefont {Khivrich}}, \bibinfo {author}
  {\bibfnamefont {J.}~\bibnamefont {Waissman}}, \bibinfo {author}
  {\bibfnamefont {K.}~\bibnamefont {Kaasbjerg}}, \bibinfo {author}
  {\bibfnamefont {Y.}~\bibnamefont {Oreg}}, \bibinfo {author} {\bibfnamefont
  {F.}~\bibnamefont {von Oppen}}, \ and\ \bibinfo {author} {\bibfnamefont
  {S.}~\bibnamefont {Ilani}},\ }\bibfield  {title} {\enquote {\bibinfo {title}
  {Electron attraction mediated by coulomb repulsion},}\ }\href {\doibase
  10.1038/nature18639} {\bibfield  {journal} {\bibinfo  {journal} {Nature}\
  }\textbf {\bibinfo {volume} {535}},\ \bibinfo {pages} {395--400} (\bibinfo
  {year} {2016})}\BibitemShut {NoStop}%
\bibitem [{\citenamefont {Delbecq}\ \emph {et~al.}(2013)\citenamefont
  {Delbecq}, \citenamefont {Bruhat}, \citenamefont {Viennot}, \citenamefont
  {Datta}, \citenamefont {Cottet},\ and\ \citenamefont
  {Kontos}}]{delbecq2013photon-mediated}%
  \BibitemOpen
  \bibfield  {author} {\bibinfo {author} {\bibfnamefont {M.~R.}\ \bibnamefont
  {Delbecq}}, \bibinfo {author} {\bibfnamefont {L.~E.}\ \bibnamefont {Bruhat}},
  \bibinfo {author} {\bibfnamefont {J.~J.}\ \bibnamefont {Viennot}}, \bibinfo
  {author} {\bibfnamefont {S.}~\bibnamefont {Datta}}, \bibinfo {author}
  {\bibfnamefont {A.}~\bibnamefont {Cottet}}, \ and\ \bibinfo {author}
  {\bibfnamefont {T.}~\bibnamefont {Kontos}},\ }\bibfield  {title} {\enquote
  {\bibinfo {title} {Photon-mediated interaction between distant quantum dot
  circuits},}\ }\href {\doibase 10.1038/ncomms2407} {\bibfield  {journal}
  {\bibinfo  {journal} {Nature Communications}\ }\textbf {\bibinfo {volume}
  {4}},\ \bibinfo {pages} {1400} (\bibinfo {year} {2013})}\BibitemShut
  {NoStop}%
\bibitem [{\citenamefont {Bhattacharya}\ \emph {et~al.}(2021)\citenamefont
  {Bhattacharya}, \citenamefont {Grass}, \citenamefont {Bachtold},
  \citenamefont {Lewenstein},\ and\ \citenamefont
  {Pistolesi}}]{bhattacharya2021phonon}%
  \BibitemOpen
  \bibfield  {author} {\bibinfo {author} {\bibfnamefont {Utso}\ \bibnamefont
  {Bhattacharya}}, \bibinfo {author} {\bibfnamefont {Tobias}\ \bibnamefont
  {Grass}}, \bibinfo {author} {\bibfnamefont {Adrian}\ \bibnamefont
  {Bachtold}}, \bibinfo {author} {\bibfnamefont {Maciej}\ \bibnamefont
  {Lewenstein}}, \ and\ \bibinfo {author} {\bibfnamefont {Fabio}\ \bibnamefont
  {Pistolesi}},\ }\bibfield  {title} {\enquote {\bibinfo {title}
  {Phonon-induced pairing in quantum dot quantum simulator},}\ }\href {\doibase
  10.1021/acs.nanolett.1c03457} {\bibfield  {journal} {\bibinfo  {journal}
  {Nano Letters}\ }\textbf {\bibinfo {volume} {21}},\ \bibinfo {pages}
  {9661--9667} (\bibinfo {year} {2021})}\BibitemShut {NoStop}%
\bibitem [{\citenamefont {Vigneau}\ \emph {et~al.}(2022)\citenamefont
  {Vigneau}, \citenamefont {Monsel}, \citenamefont {Tabanera}, \citenamefont
  {Aggarwal}, \citenamefont {Bresque}, \citenamefont {Fedele}, \citenamefont
  {Cerisola}, \citenamefont {Briggs}, \citenamefont {Anders}, \citenamefont
  {Parrondo}, \citenamefont {Auff\`eves},\ and\ \citenamefont
  {Ares}}]{vigneauPRRultrastrong}%
  \BibitemOpen
  \bibfield  {author} {\bibinfo {author} {\bibfnamefont {Florian}\ \bibnamefont
  {Vigneau}}, \bibinfo {author} {\bibfnamefont {Juliette}\ \bibnamefont
  {Monsel}}, \bibinfo {author} {\bibfnamefont {Jorge}\ \bibnamefont
  {Tabanera}}, \bibinfo {author} {\bibfnamefont {Kushagra}\ \bibnamefont
  {Aggarwal}}, \bibinfo {author} {\bibfnamefont {L\'ea}\ \bibnamefont
  {Bresque}}, \bibinfo {author} {\bibfnamefont {Federico}\ \bibnamefont
  {Fedele}}, \bibinfo {author} {\bibfnamefont {Federico}\ \bibnamefont
  {Cerisola}}, \bibinfo {author} {\bibfnamefont {G.~A.~D.}\ \bibnamefont
  {Briggs}}, \bibinfo {author} {\bibfnamefont {Janet}\ \bibnamefont {Anders}},
  \bibinfo {author} {\bibfnamefont {Juan M.~R.}\ \bibnamefont {Parrondo}},
  \bibinfo {author} {\bibfnamefont {Alexia}\ \bibnamefont {Auff\`eves}}, \ and\
  \bibinfo {author} {\bibfnamefont {Natalia}\ \bibnamefont {Ares}},\ }\bibfield
   {title} {\enquote {\bibinfo {title} {Ultrastrong coupling between electron
  tunneling and mechanical motion},}\ }\href {\doibase
  10.1103/PhysRevResearch.4.043168} {\bibfield  {journal} {\bibinfo  {journal}
  {Phys. Rev. Res.}\ }\textbf {\bibinfo {volume} {4}},\ \bibinfo {pages}
  {043168} (\bibinfo {year} {2022})}\BibitemShut {NoStop}%
\bibitem [{\citenamefont {Moser}\ \emph {et~al.}(2013)\citenamefont {Moser},
  \citenamefont {G{\"u}ttinger}, \citenamefont {Eichler}, \citenamefont
  {Esplandiu}, \citenamefont {Liu}, \citenamefont {Dykman},\ and\ \citenamefont
  {Bachtold}}]{moser2013ultrasensitive}%
  \BibitemOpen
  \bibfield  {author} {\bibinfo {author} {\bibfnamefont {J.}~\bibnamefont
  {Moser}}, \bibinfo {author} {\bibfnamefont {J.}~\bibnamefont
  {G{\"u}ttinger}}, \bibinfo {author} {\bibfnamefont {A.}~\bibnamefont
  {Eichler}}, \bibinfo {author} {\bibfnamefont {M.~J.}\ \bibnamefont
  {Esplandiu}}, \bibinfo {author} {\bibfnamefont {D.~E.}\ \bibnamefont {Liu}},
  \bibinfo {author} {\bibfnamefont {M.~I.}\ \bibnamefont {Dykman}}, \ and\
  \bibinfo {author} {\bibfnamefont {A.}~\bibnamefont {Bachtold}},\ }\bibfield
  {title} {\enquote {\bibinfo {title} {Ultrasensitive force detection with a
  nanotube mechanical resonator},}\ }\href {\doibase 10.1038/nnano.2013.97}
  {\bibfield  {journal} {\bibinfo  {journal} {Nature Nanotechnology}\ }\textbf
  {\bibinfo {volume} {8}},\ \bibinfo {pages} {493--496} (\bibinfo {year}
  {2013})}\BibitemShut {NoStop}%
\bibitem [{\citenamefont {Riwar}\ \emph {et~al.}(2016)\citenamefont {Riwar},
  \citenamefont {Houzet}, \citenamefont {Meyer},\ and\ \citenamefont
  {Nazarov}}]{riwar2016multi-terminal}%
  \BibitemOpen
  \bibfield  {author} {\bibinfo {author} {\bibfnamefont {Roman-Pascal}\
  \bibnamefont {Riwar}}, \bibinfo {author} {\bibfnamefont {Manuel}\
  \bibnamefont {Houzet}}, \bibinfo {author} {\bibfnamefont {Julia~S.}\
  \bibnamefont {Meyer}}, \ and\ \bibinfo {author} {\bibfnamefont {Yuli~V.}\
  \bibnamefont {Nazarov}},\ }\bibfield  {title} {\enquote {\bibinfo {title}
  {Multi-terminal josephson junctions as topological matter},}\ }\href
  {\doibase 10.1038/ncomms11167} {\bibfield  {journal} {\bibinfo  {journal}
  {Nature Communications}\ }\textbf {\bibinfo {volume} {7}},\ \bibinfo {pages}
  {11167} (\bibinfo {year} {2016})}\BibitemShut {NoStop}%
\bibitem [{\citenamefont {Teshler}\ \emph {et~al.}(2023)\citenamefont
  {Teshler}, \citenamefont {Weisbrich}, \citenamefont {Sturm}, \citenamefont
  {Klees}, \citenamefont {Rastelli},\ and\ \citenamefont
  {Belzig}}]{teshler2023ground}%
  \BibitemOpen
  \bibfield  {author} {\bibinfo {author} {\bibfnamefont {Lev}\ \bibnamefont
  {Teshler}}, \bibinfo {author} {\bibfnamefont {Hannes}\ \bibnamefont
  {Weisbrich}}, \bibinfo {author} {\bibfnamefont {Jonathan}\ \bibnamefont
  {Sturm}}, \bibinfo {author} {\bibfnamefont {Raffael~L.}\ \bibnamefont
  {Klees}}, \bibinfo {author} {\bibfnamefont {Gianluca}\ \bibnamefont
  {Rastelli}}, \ and\ \bibinfo {author} {\bibfnamefont {Wolfgang}\ \bibnamefont
  {Belzig}},\ }\bibfield  {title} {\enquote {\bibinfo {title} {{Ground state
  topology of a four-terminal superconducting double quantum dot}},}\ }\href
  {\doibase 10.21468/SciPostPhys.15.5.214} {\bibfield  {journal} {\bibinfo
  {journal} {SciPost Phys.}\ }\textbf {\bibinfo {volume} {15}},\ \bibinfo
  {pages} {214} (\bibinfo {year} {2023})}\BibitemShut {NoStop}%
\bibitem [{\citenamefont {Ohnmacht}\ \emph {et~al.}(2024)\citenamefont
  {Ohnmacht}, \citenamefont {Coraiola}, \citenamefont {Garc\'{\i}a-Esteban},
  \citenamefont {Sabonis}, \citenamefont {Nichele}, \citenamefont {Belzig},\
  and\ \citenamefont {Cuevas}}]{ohnmacht2023quartet}%
  \BibitemOpen
  \bibfield  {author} {\bibinfo {author} {\bibfnamefont {David~Christian}\
  \bibnamefont {Ohnmacht}}, \bibinfo {author} {\bibfnamefont {Marco}\
  \bibnamefont {Coraiola}}, \bibinfo {author} {\bibfnamefont {Juan~Jos\'e}\
  \bibnamefont {Garc\'{\i}a-Esteban}}, \bibinfo {author} {\bibfnamefont
  {Deividas}\ \bibnamefont {Sabonis}}, \bibinfo {author} {\bibfnamefont
  {Fabrizio}\ \bibnamefont {Nichele}}, \bibinfo {author} {\bibfnamefont
  {Wolfgang}\ \bibnamefont {Belzig}}, \ and\ \bibinfo {author} {\bibfnamefont
  {Juan~Carlos}\ \bibnamefont {Cuevas}},\ }\bibfield  {title} {\enquote
  {\bibinfo {title} {Quartet tomography in multiterminal josephson
  junctions},}\ }\href {\doibase 10.1103/PhysRevB.109.L241407} {\bibfield
  {journal} {\bibinfo  {journal} {Phys. Rev. B}\ }\textbf {\bibinfo {volume}
  {109}},\ \bibinfo {pages} {L241407} (\bibinfo {year} {2024})}\BibitemShut
  {NoStop}%
\bibitem [{\citenamefont {Zalom}\ \emph {et~al.}(2024)\citenamefont {Zalom},
  \citenamefont {\ifmmode~\check{Z}\else \v{Z}\fi{}onda},\ and\ \citenamefont
  {Novotn\'y}}]{zalom2024hidden}%
  \BibitemOpen
  \bibfield  {author} {\bibinfo {author} {\bibfnamefont {Peter}\ \bibnamefont
  {Zalom}}, \bibinfo {author} {\bibfnamefont {M.}~\bibnamefont
  {\ifmmode~\check{Z}\else \v{Z}\fi{}onda}}, \ and\ \bibinfo {author}
  {\bibfnamefont {T.}~\bibnamefont {Novotn\'y}},\ }\bibfield  {title} {\enquote
  {\bibinfo {title} {Hidden symmetry in interacting-quantum-dot-based
  multiterminal josephson junctions},}\ }\href {\doibase
  10.1103/PhysRevLett.132.126505} {\bibfield  {journal} {\bibinfo  {journal}
  {Phys. Rev. Lett.}\ }\textbf {\bibinfo {volume} {132}},\ \bibinfo {pages}
  {126505} (\bibinfo {year} {2024})}\BibitemShut {NoStop}%
\bibitem [{\citenamefont {Coraiola}\ \emph {et~al.}(2023)\citenamefont
  {Coraiola}, \citenamefont {Haxell}, \citenamefont {Sabonis}, \citenamefont
  {Weisbrich}, \citenamefont {Svetogorov}, \citenamefont {Hinderling},
  \citenamefont {ten Kate}, \citenamefont {Cheah}, \citenamefont {Krizek},
  \citenamefont {Schott}, \citenamefont {Wegscheider}, \citenamefont {Cuevas},
  \citenamefont {Belzig},\ and\ \citenamefont
  {Nichele}}]{coraiola2023phase-engineering}%
  \BibitemOpen
  \bibfield  {author} {\bibinfo {author} {\bibfnamefont {Marco}\ \bibnamefont
  {Coraiola}}, \bibinfo {author} {\bibfnamefont {Daniel~Z.}\ \bibnamefont
  {Haxell}}, \bibinfo {author} {\bibfnamefont {Deividas}\ \bibnamefont
  {Sabonis}}, \bibinfo {author} {\bibfnamefont {Hannes}\ \bibnamefont
  {Weisbrich}}, \bibinfo {author} {\bibfnamefont {Aleksandr~E.}\ \bibnamefont
  {Svetogorov}}, \bibinfo {author} {\bibfnamefont {Manuel}\ \bibnamefont
  {Hinderling}}, \bibinfo {author} {\bibfnamefont {Sofieke~C.}\ \bibnamefont
  {ten Kate}}, \bibinfo {author} {\bibfnamefont {Erik}\ \bibnamefont {Cheah}},
  \bibinfo {author} {\bibfnamefont {Filip}\ \bibnamefont {Krizek}}, \bibinfo
  {author} {\bibfnamefont {R{\"u}diger}\ \bibnamefont {Schott}}, \bibinfo
  {author} {\bibfnamefont {Werner}\ \bibnamefont {Wegscheider}}, \bibinfo
  {author} {\bibfnamefont {Juan~Carlos}\ \bibnamefont {Cuevas}}, \bibinfo
  {author} {\bibfnamefont {Wolfgang}\ \bibnamefont {Belzig}}, \ and\ \bibinfo
  {author} {\bibfnamefont {Fabrizio}\ \bibnamefont {Nichele}},\ }\bibfield
  {title} {\enquote {\bibinfo {title} {Phase-engineering the andreev band
  structure of a three-terminal josephson junction},}\ }\href {\doibase
  10.1038/s41467-023-42356-6} {\bibfield  {journal} {\bibinfo  {journal}
  {Nature Communications}\ }\textbf {\bibinfo {volume} {14}},\ \bibinfo {pages}
  {6784} (\bibinfo {year} {2023})}\BibitemShut {NoStop}%
\bibitem [{\citenamefont {Pankratova}\ \emph {et~al.}(2020)\citenamefont
  {Pankratova}, \citenamefont {Lee}, \citenamefont {Kuzmin}, \citenamefont
  {Wickramasinghe}, \citenamefont {Mayer}, \citenamefont {Yuan}, \citenamefont
  {Vavilov}, \citenamefont {Shabani},\ and\ \citenamefont
  {Manucharyan}}]{pankratova2020multiterminal}%
  \BibitemOpen
  \bibfield  {author} {\bibinfo {author} {\bibfnamefont {Natalia}\ \bibnamefont
  {Pankratova}}, \bibinfo {author} {\bibfnamefont {Hanho}\ \bibnamefont {Lee}},
  \bibinfo {author} {\bibfnamefont {Roman}\ \bibnamefont {Kuzmin}}, \bibinfo
  {author} {\bibfnamefont {Kaushini}\ \bibnamefont {Wickramasinghe}}, \bibinfo
  {author} {\bibfnamefont {William}\ \bibnamefont {Mayer}}, \bibinfo {author}
  {\bibfnamefont {Joseph}\ \bibnamefont {Yuan}}, \bibinfo {author}
  {\bibfnamefont {Maxim~G.}\ \bibnamefont {Vavilov}}, \bibinfo {author}
  {\bibfnamefont {Javad}\ \bibnamefont {Shabani}}, \ and\ \bibinfo {author}
  {\bibfnamefont {Vladimir~E.}\ \bibnamefont {Manucharyan}},\ }\bibfield
  {title} {\enquote {\bibinfo {title} {Multiterminal josephson effect},}\
  }\href {\doibase 10.1103/PhysRevX.10.031051} {\bibfield  {journal} {\bibinfo
  {journal} {Phys. Rev. X}\ }\textbf {\bibinfo {volume} {10}},\ \bibinfo
  {pages} {031051} (\bibinfo {year} {2020})}\BibitemShut {NoStop}%
\bibitem [{\citenamefont {Matsuo}\ \emph {et~al.}(2022)\citenamefont {Matsuo},
  \citenamefont {Lee}, \citenamefont {Chang}, \citenamefont {Sato},
  \citenamefont {Ueda}, \citenamefont {Palmstr{\o}m},\ and\ \citenamefont
  {Tarucha}}]{matsuo2022observation}%
  \BibitemOpen
  \bibfield  {author} {\bibinfo {author} {\bibfnamefont {Sadashige}\
  \bibnamefont {Matsuo}}, \bibinfo {author} {\bibfnamefont {Joon~Sue}\
  \bibnamefont {Lee}}, \bibinfo {author} {\bibfnamefont {Chien-Yuan}\
  \bibnamefont {Chang}}, \bibinfo {author} {\bibfnamefont {Yosuke}\
  \bibnamefont {Sato}}, \bibinfo {author} {\bibfnamefont {Kento}\ \bibnamefont
  {Ueda}}, \bibinfo {author} {\bibfnamefont {Christopher~J.}\ \bibnamefont
  {Palmstr{\o}m}}, \ and\ \bibinfo {author} {\bibfnamefont {Seigo}\
  \bibnamefont {Tarucha}},\ }\bibfield  {title} {\enquote {\bibinfo {title}
  {Observation of nonlocal josephson effect on double inas nanowires},}\ }\href
  {\doibase 10.1038/s42005-022-00994-0} {\bibfield  {journal} {\bibinfo
  {journal} {Communications Physics}\ }\textbf {\bibinfo {volume} {5}},\
  \bibinfo {pages} {221} (\bibinfo {year} {2022})}\BibitemShut {NoStop}%
\bibitem [{\citenamefont {Blatter}\ \emph {et~al.}(2001)\citenamefont
  {Blatter}, \citenamefont {Geshkenbein},\ and\ \citenamefont
  {Ioffe}}]{blatter2001design}%
  \BibitemOpen
  \bibfield  {author} {\bibinfo {author} {\bibfnamefont {Gianni}\ \bibnamefont
  {Blatter}}, \bibinfo {author} {\bibfnamefont {Vadim~B.}\ \bibnamefont
  {Geshkenbein}}, \ and\ \bibinfo {author} {\bibfnamefont {Lev~B.}\
  \bibnamefont {Ioffe}},\ }\bibfield  {title} {\enquote {\bibinfo {title}
  {Design aspects of superconducting-phase quantum bits},}\ }\href {\doibase
  10.1103/PhysRevB.63.174511} {\bibfield  {journal} {\bibinfo  {journal} {Phys.
  Rev. B}\ }\textbf {\bibinfo {volume} {63}},\ \bibinfo {pages} {174511}
  (\bibinfo {year} {2001})}\BibitemShut {NoStop}%
\bibitem [{\citenamefont {Ioffe}\ \emph {et~al.}(2002)\citenamefont {Ioffe},
  \citenamefont {Feigel'man}, \citenamefont {Ioselevich}, \citenamefont
  {Ivanov}, \citenamefont {Troyer},\ and\ \citenamefont
  {Blatter}}]{Ioffe2002topologically}%
  \BibitemOpen
  \bibfield  {author} {\bibinfo {author} {\bibfnamefont {L.~B.}\ \bibnamefont
  {Ioffe}}, \bibinfo {author} {\bibfnamefont {M.~V.}\ \bibnamefont
  {Feigel'man}}, \bibinfo {author} {\bibfnamefont {A.}~\bibnamefont
  {Ioselevich}}, \bibinfo {author} {\bibfnamefont {D.}~\bibnamefont {Ivanov}},
  \bibinfo {author} {\bibfnamefont {M.}~\bibnamefont {Troyer}}, \ and\ \bibinfo
  {author} {\bibfnamefont {G.}~\bibnamefont {Blatter}},\ }\bibfield  {title}
  {\enquote {\bibinfo {title} {Topologically protected quantum bits using
  josephson junction arrays},}\ }\href {\doibase 10.1038/415503a} {\bibfield
  {journal} {\bibinfo  {journal} {Nature}\ }\textbf {\bibinfo {volume} {415}},\
  \bibinfo {pages} {503--506} (\bibinfo {year} {2002})}\BibitemShut {NoStop}%
\bibitem [{\citenamefont {Ioffe}\ and\ \citenamefont
  {Feigel'man}(2002)}]{ioffe2002possible}%
  \BibitemOpen
  \bibfield  {author} {\bibinfo {author} {\bibfnamefont {L.~B.}\ \bibnamefont
  {Ioffe}}\ and\ \bibinfo {author} {\bibfnamefont {M.~V.}\ \bibnamefont
  {Feigel'man}},\ }\bibfield  {title} {\enquote {\bibinfo {title} {Possible
  realization of an ideal quantum computer in josephson junction array},}\
  }\href {\doibase 10.1103/PhysRevB.66.224503} {\bibfield  {journal} {\bibinfo
  {journal} {Phys. Rev. B}\ }\textbf {\bibinfo {volume} {66}},\ \bibinfo
  {pages} {224503} (\bibinfo {year} {2002})}\BibitemShut {NoStop}%
\bibitem [{\citenamefont {Brooks}\ \emph {et~al.}(2013)\citenamefont {Brooks},
  \citenamefont {Kitaev},\ and\ \citenamefont
  {Preskill}}]{brooks2013protected}%
  \BibitemOpen
  \bibfield  {author} {\bibinfo {author} {\bibfnamefont {Peter}\ \bibnamefont
  {Brooks}}, \bibinfo {author} {\bibfnamefont {Alexei}\ \bibnamefont {Kitaev}},
  \ and\ \bibinfo {author} {\bibfnamefont {John}\ \bibnamefont {Preskill}},\
  }\bibfield  {title} {\enquote {\bibinfo {title} {Protected gates for
  superconducting qubits},}\ }\href {\doibase 10.1103/PhysRevA.87.052306}
  {\bibfield  {journal} {\bibinfo  {journal} {Phys. Rev. A}\ }\textbf {\bibinfo
  {volume} {87}},\ \bibinfo {pages} {052306} (\bibinfo {year}
  {2013})}\BibitemShut {NoStop}%
\bibitem [{\citenamefont {Dou{\c c}ot}\ and\ \citenamefont
  {Ioffe}(2012)}]{doucot2012physical}%
  \BibitemOpen
  \bibfield  {author} {\bibinfo {author} {\bibfnamefont {B}~\bibnamefont
  {Dou{\c c}ot}}\ and\ \bibinfo {author} {\bibfnamefont {L~B}\ \bibnamefont
  {Ioffe}},\ }\bibfield  {title} {\enquote {\bibinfo {title} {Physical
  implementation of protected qubits},}\ }\href {\doibase
  10.1088/0034-4885/75/7/072001} {\bibfield  {journal} {\bibinfo  {journal}
  {Reports on Progress in Physics}\ }\textbf {\bibinfo {volume} {75}},\
  \bibinfo {pages} {072001} (\bibinfo {year} {2012})}\BibitemShut {NoStop}%
\bibitem [{\citenamefont {Brosco}\ \emph {et~al.}(2024)\citenamefont {Brosco},
  \citenamefont {Serpico}, \citenamefont {Vinokur}, \citenamefont {Poccia},\
  and\ \citenamefont {Vool}}]{brosco2024superconducting}%
  \BibitemOpen
  \bibfield  {author} {\bibinfo {author} {\bibfnamefont {Valentina}\
  \bibnamefont {Brosco}}, \bibinfo {author} {\bibfnamefont {Giuseppe}\
  \bibnamefont {Serpico}}, \bibinfo {author} {\bibfnamefont {Valerii}\
  \bibnamefont {Vinokur}}, \bibinfo {author} {\bibfnamefont {Nicola}\
  \bibnamefont {Poccia}}, \ and\ \bibinfo {author} {\bibfnamefont {Uri}\
  \bibnamefont {Vool}},\ }\bibfield  {title} {\enquote {\bibinfo {title}
  {Superconducting qubit based on twisted cuprate van der waals
  heterostructures},}\ }\href {\doibase 10.1103/PhysRevLett.132.017003}
  {\bibfield  {journal} {\bibinfo  {journal} {Phys. Rev. Lett.}\ }\textbf
  {\bibinfo {volume} {132}},\ \bibinfo {pages} {017003} (\bibinfo {year}
  {2024})}\BibitemShut {NoStop}%
\bibitem [{\citenamefont {Coppo}\ \emph {et~al.}(2024)\citenamefont {Coppo},
  \citenamefont {Chirolli}, \citenamefont {Poccia}, \citenamefont {Vool},\ and\
  \citenamefont {Brosco}}]{coppo2024flux-tunable}%
  \BibitemOpen
  \bibfield  {author} {\bibinfo {author} {\bibfnamefont {Alessandro}\
  \bibnamefont {Coppo}}, \bibinfo {author} {\bibfnamefont {Luca}\ \bibnamefont
  {Chirolli}}, \bibinfo {author} {\bibfnamefont {Nicola}\ \bibnamefont
  {Poccia}}, \bibinfo {author} {\bibfnamefont {Uri}\ \bibnamefont {Vool}}, \
  and\ \bibinfo {author} {\bibfnamefont {Valentina}\ \bibnamefont {Brosco}},\
  }\bibfield  {title} {\enquote {\bibinfo {title} {{Flux-tunable regimes and
  supersymmetry in twisted cuprate heterostructures}},}\ }\href {\doibase
  10.1063/5.0217614} {\bibfield  {journal} {\bibinfo  {journal} {Applied
  Physics Letters}\ }\textbf {\bibinfo {volume} {125}},\ \bibinfo {pages}
  {054001} (\bibinfo {year} {2024})},\ \Eprint
  {http://arxiv.org/abs/https://pubs.aip.org/aip/apl/article-pdf/doi/10.1063/5.0217614/20084254/054001\_1\_5.0217614.pdf}
  {https://pubs.aip.org/aip/apl/article-pdf/doi/10.1063/5.0217614/20084254/054001\_1\_5.0217614.pdf}
  \BibitemShut {NoStop}%
\bibitem [{\citenamefont {Ryazanov}\ \emph {et~al.}(2001)\citenamefont
  {Ryazanov}, \citenamefont {Oboznov}, \citenamefont {Rusanov}, \citenamefont
  {Veretennikov}, \citenamefont {Golubov},\ and\ \citenamefont
  {Aarts}}]{golubov2001coupling}%
  \BibitemOpen
  \bibfield  {author} {\bibinfo {author} {\bibfnamefont {V.~V.}\ \bibnamefont
  {Ryazanov}}, \bibinfo {author} {\bibfnamefont {V.~A.}\ \bibnamefont
  {Oboznov}}, \bibinfo {author} {\bibfnamefont {A.~Yu.}\ \bibnamefont
  {Rusanov}}, \bibinfo {author} {\bibfnamefont {A.~V.}\ \bibnamefont
  {Veretennikov}}, \bibinfo {author} {\bibfnamefont {A.~A.}\ \bibnamefont
  {Golubov}}, \ and\ \bibinfo {author} {\bibfnamefont {J.}~\bibnamefont
  {Aarts}},\ }\bibfield  {title} {\enquote {\bibinfo {title} {Coupling of two
  superconductors through a ferromagnet: Evidence for a $\ensuremath{\pi}$
  junction},}\ }\href {\doibase 10.1103/PhysRevLett.86.2427} {\bibfield
  {journal} {\bibinfo  {journal} {Phys. Rev. Lett.}\ }\textbf {\bibinfo
  {volume} {86}},\ \bibinfo {pages} {2427--2430} (\bibinfo {year}
  {2001})}\BibitemShut {NoStop}%
\bibitem [{\citenamefont {Tsuei}\ and\ \citenamefont
  {Kirtley}(2000)}]{tsuei2000pairing}%
  \BibitemOpen
  \bibfield  {author} {\bibinfo {author} {\bibfnamefont {C.~C.}\ \bibnamefont
  {Tsuei}}\ and\ \bibinfo {author} {\bibfnamefont {J.~R.}\ \bibnamefont
  {Kirtley}},\ }\bibfield  {title} {\enquote {\bibinfo {title} {Pairing
  symmetry in cuprate superconductors},}\ }\href {\doibase
  10.1103/RevModPhys.72.969} {\bibfield  {journal} {\bibinfo  {journal} {Rev.
  Mod. Phys.}\ }\textbf {\bibinfo {volume} {72}},\ \bibinfo {pages} {969--1016}
  (\bibinfo {year} {2000})}\BibitemShut {NoStop}%
\bibitem [{\citenamefont {Baselmans}\ \emph {et~al.}(1999)\citenamefont
  {Baselmans}, \citenamefont {Morpurgo}, \citenamefont {van Wees},\ and\
  \citenamefont {Klapwijk}}]{baselmans1999reversing}%
  \BibitemOpen
  \bibfield  {author} {\bibinfo {author} {\bibfnamefont {J.~J.~A.}\
  \bibnamefont {Baselmans}}, \bibinfo {author} {\bibfnamefont {A.~F.}\
  \bibnamefont {Morpurgo}}, \bibinfo {author} {\bibfnamefont {B.~J.}\
  \bibnamefont {van Wees}}, \ and\ \bibinfo {author} {\bibfnamefont {T.~M.}\
  \bibnamefont {Klapwijk}},\ }\bibfield  {title} {\enquote {\bibinfo {title}
  {Reversing the direction of the supercurrent in a controllable josephson
  junction},}\ }\href {\doibase 10.1038/16204} {\bibfield  {journal} {\bibinfo
  {journal} {Nature}\ }\textbf {\bibinfo {volume} {397}},\ \bibinfo {pages}
  {43--45} (\bibinfo {year} {1999})}\BibitemShut {NoStop}%
\bibitem [{\citenamefont {Cleuziou}\ \emph {et~al.}(2006)\citenamefont
  {Cleuziou}, \citenamefont {Wernsdorfer}, \citenamefont {Bouchiat},
  \citenamefont {Ondar{\c c}uhu},\ and\ \citenamefont
  {Monthioux}}]{cleuziou2006carbon}%
  \BibitemOpen
  \bibfield  {author} {\bibinfo {author} {\bibfnamefont {J.~P.}\ \bibnamefont
  {Cleuziou}}, \bibinfo {author} {\bibfnamefont {W.}~\bibnamefont
  {Wernsdorfer}}, \bibinfo {author} {\bibfnamefont {V.}~\bibnamefont
  {Bouchiat}}, \bibinfo {author} {\bibfnamefont {T.}~\bibnamefont {Ondar{\c
  c}uhu}}, \ and\ \bibinfo {author} {\bibfnamefont {M.}~\bibnamefont
  {Monthioux}},\ }\bibfield  {title} {\enquote {\bibinfo {title} {Carbon
  nanotube superconducting quantum interference device},}\ }\href {\doibase
  10.1038/nnano.2006.54} {\bibfield  {journal} {\bibinfo  {journal} {Nature
  Nanotechnology}\ }\textbf {\bibinfo {volume} {1}},\ \bibinfo {pages} {53--59}
  (\bibinfo {year} {2006})}\BibitemShut {NoStop}%
\bibitem [{\citenamefont {Szombati}\ \emph {et~al.}(2016)\citenamefont
  {Szombati}, \citenamefont {Nadj-Perge}, \citenamefont {Car}, \citenamefont
  {Plissard}, \citenamefont {Bakkers},\ and\ \citenamefont
  {Kouwenhoven}}]{szombati2016josephson}%
  \BibitemOpen
  \bibfield  {author} {\bibinfo {author} {\bibfnamefont {D.~B.}\ \bibnamefont
  {Szombati}}, \bibinfo {author} {\bibfnamefont {S.}~\bibnamefont
  {Nadj-Perge}}, \bibinfo {author} {\bibfnamefont {D.}~\bibnamefont {Car}},
  \bibinfo {author} {\bibfnamefont {S.~R.}\ \bibnamefont {Plissard}}, \bibinfo
  {author} {\bibfnamefont {E.~P. A.~M.}\ \bibnamefont {Bakkers}}, \ and\
  \bibinfo {author} {\bibfnamefont {L.~P.}\ \bibnamefont {Kouwenhoven}},\
  }\bibfield  {title} {\enquote {\bibinfo {title} {Josephson $\phi_0$-junction
  in nanowire quantum dots},}\ }\href {\doibase 10.1038/nphys3742} {\bibfield
  {journal} {\bibinfo  {journal} {Nature Physics}\ }\textbf {\bibinfo {volume}
  {12}},\ \bibinfo {pages} {568--572} (\bibinfo {year} {2016})}\BibitemShut
  {NoStop}%
\bibitem [{\citenamefont {Mahan}\ and\ \citenamefont {Jeon}(2004)}]{mahan2004}%
  \BibitemOpen
  \bibfield  {author} {\bibinfo {author} {\bibfnamefont {G.~D.}\ \bibnamefont
  {Mahan}}\ and\ \bibinfo {author} {\bibfnamefont {Gun~Sang}\ \bibnamefont
  {Jeon}},\ }\bibfield  {title} {\enquote {\bibinfo {title} {Flexure modes in
  carbon nanotubes},}\ }\href {\doibase 10.1103/PhysRevB.70.075405} {\bibfield
  {journal} {\bibinfo  {journal} {Phys. Rev. B}\ }\textbf {\bibinfo {volume}
  {70}},\ \bibinfo {pages} {075405} (\bibinfo {year} {2004})}\BibitemShut
  {NoStop}%
\bibitem [{\citenamefont {Biercuk}\ \emph {et~al.}(2005)\citenamefont
  {Biercuk}, \citenamefont {Garaj}, \citenamefont {Mason}, \citenamefont
  {Chow},\ and\ \citenamefont {Marcus}}]{biercuk2005gate-defined}%
  \BibitemOpen
  \bibfield  {author} {\bibinfo {author} {\bibfnamefont {M.~J.}\ \bibnamefont
  {Biercuk}}, \bibinfo {author} {\bibfnamefont {S.}~\bibnamefont {Garaj}},
  \bibinfo {author} {\bibfnamefont {N.}~\bibnamefont {Mason}}, \bibinfo
  {author} {\bibfnamefont {J.~M.}\ \bibnamefont {Chow}}, \ and\ \bibinfo
  {author} {\bibfnamefont {C.~M.}\ \bibnamefont {Marcus}},\ }\bibfield  {title}
  {\enquote {\bibinfo {title} {Gate-defined quantum dots on carbon
  nanotubes},}\ }\href {\doibase 10.1021/nl050364v} {\bibfield  {journal}
  {\bibinfo  {journal} {Nano Letters}\ }\textbf {\bibinfo {volume} {5}},\
  \bibinfo {pages} {1267--1271} (\bibinfo {year} {2005})}\BibitemShut {NoStop}%
\bibitem [{\citenamefont {Hofstetter}\ \emph {et~al.}(2009)\citenamefont
  {Hofstetter}, \citenamefont {Csonka}, \citenamefont {Nyg{\aa}rd},\ and\
  \citenamefont {Sch{\"o}nenberger}}]{hofstetter2009cooper}%
  \BibitemOpen
  \bibfield  {author} {\bibinfo {author} {\bibfnamefont {L.}~\bibnamefont
  {Hofstetter}}, \bibinfo {author} {\bibfnamefont {S.}~\bibnamefont {Csonka}},
  \bibinfo {author} {\bibfnamefont {J.}~\bibnamefont {Nyg{\aa}rd}}, \ and\
  \bibinfo {author} {\bibfnamefont {C.}~\bibnamefont {Sch{\"o}nenberger}},\
  }\bibfield  {title} {\enquote {\bibinfo {title} {Cooper pair splitter
  realized in a two-quantum-dot y-junction},}\ }\href {\doibase
  10.1038/nature08432} {\bibfield  {journal} {\bibinfo  {journal} {Nature}\
  }\textbf {\bibinfo {volume} {461}},\ \bibinfo {pages} {960--963} (\bibinfo
  {year} {2009})}\BibitemShut {NoStop}%
\bibitem [{\citenamefont {Herrmann}\ \emph {et~al.}(2010)\citenamefont
  {Herrmann}, \citenamefont {Portier}, \citenamefont {Roche}, \citenamefont
  {Yeyati}, \citenamefont {Kontos},\ and\ \citenamefont
  {Strunk}}]{herrmann2010carbon}%
  \BibitemOpen
  \bibfield  {author} {\bibinfo {author} {\bibfnamefont {L.~G.}\ \bibnamefont
  {Herrmann}}, \bibinfo {author} {\bibfnamefont {F.}~\bibnamefont {Portier}},
  \bibinfo {author} {\bibfnamefont {P.}~\bibnamefont {Roche}}, \bibinfo
  {author} {\bibfnamefont {A.~Levy}\ \bibnamefont {Yeyati}}, \bibinfo {author}
  {\bibfnamefont {T.}~\bibnamefont {Kontos}}, \ and\ \bibinfo {author}
  {\bibfnamefont {C.}~\bibnamefont {Strunk}},\ }\bibfield  {title} {\enquote
  {\bibinfo {title} {Carbon nanotubes as cooper-pair beam splitters},}\ }\href
  {\doibase 10.1103/PhysRevLett.104.026801} {\bibfield  {journal} {\bibinfo
  {journal} {Phys. Rev. Lett.}\ }\textbf {\bibinfo {volume} {104}},\ \bibinfo
  {pages} {026801} (\bibinfo {year} {2010})}\BibitemShut {NoStop}%
\bibitem [{\citenamefont {Soldini}\ \emph {et~al.}(2024)\citenamefont
  {Soldini}, \citenamefont {Fischer},\ and\ \citenamefont
  {Neupert}}]{soldini2024charge4e}%
  \BibitemOpen
  \bibfield  {author} {\bibinfo {author} {\bibfnamefont {Martina~O.}\
  \bibnamefont {Soldini}}, \bibinfo {author} {\bibfnamefont {Mark~H.}\
  \bibnamefont {Fischer}}, \ and\ \bibinfo {author} {\bibfnamefont {Titus}\
  \bibnamefont {Neupert}},\ }\bibfield  {title} {\enquote {\bibinfo {title}
  {Charge-$4e$ superconductivity in a hubbard model},}\ }\href {\doibase
  10.1103/PhysRevB.109.214509} {\bibfield  {journal} {\bibinfo  {journal}
  {Phys. Rev. B}\ }\textbf {\bibinfo {volume} {109}},\ \bibinfo {pages}
  {214509} (\bibinfo {year} {2024})}\BibitemShut {NoStop}%
\end{thebibliography}%

\end{document}